\documentclass[%
 aip,
 amsmath,amssymb,
 reprint,%
]{revtex4-2}

\usepackage{graphicx}
\usepackage{dcolumn}
\usepackage{bm}
\usepackage{color}
\usepackage[T1]{fontenc}
\usepackage{mathptmx}
\usepackage{etoolbox}

\makeatletter
\def\@email#1#2{%
 \endgroup
 \patchcmd{\titleblock@produce}
  {\frontmatter@RRAPformat}
  {\frontmatter@RRAPformat{\produce@RRAP{*#1\href{mailto:#2}{#2}}}\frontmatter@RRAPformat}
  {}{}
}%
\makeatother
\begin{document}

\preprint{}

\title[Quantum vortex identification method and its application to Gross-Pitaevskii simulation]
{Quantum vortex identification method and its application to Gross-Pitaevskii simulation}
\author{Naoto Sakaki}
\email{sakaki.nu.2023@gmail.com}
\affiliation{ 
Graduate School of Engineering, Nagoya University, Furocho, Chikusa, Nagoya 464-8603, Japan 
}%

\author{Hideaki Miura}%
\email{miura.hideaki@nifs.ac.jp}
\affiliation{ 
Complex Global Simulation Unit, National Institute for Fusion Science, Toki, Gifu 509-5292, Japan %
}%

\author{Kyo Yoshida}
\affiliation{%
Faculty of Pure and Applied Sciences, University of Tsukuba, 1-1-1 Tennoudai, Tsukuba, Ibaraki 305-8571, Japan
}%

\author{Yoshiyuki Tsuji}
\affiliation{%
Graduate School of Engineering, Nagoya University, Furocho, Chikusa, Nagoya 464-8603, Japan
}%

\date{\today}

\begin{abstract}
A method to identify a quantum vortex in a three-dimensional Gross-Pitaevskii simulation has been developed.
A quantum vortex was identified by the use of eigenvalues and eigenvectors of the Hessian of the mass density, together with a condition to distinguish a point to constitute a swirling vortex from other confusing data points.
This method has been verified to identify vortex axes in a Gross-Pitaevskii simulation appropriately, being useful to elucidate various statistics associated with turbulent quantum vortices.
This method provides us with a unified approach to studying vortex statistics in turbulence of both classic and quantum fluids.
Our study reveals that the maximum radius of a swirling region of a quantum vortex can be as large as sixty times the healing length.
The characterization of the vortex core radius relative to the healing length is reported for the first time in this paper.
Furthermore, the geometrical natures of vortex axes such as the probability density function of the curvature are characterized by the healing length.
\end{abstract}

\maketitle

\section{\label{sec:sec1}Introduction}
Quantum vortices in a superfluid get attention because they represent intrinsic feature of fluid motion, the inviscid limit of classic turbulence.
(See reviews in Refs.\cite{Vinen2002,Tsubota2008,Paoletti2011} and references therein, for example.)
Quantum vortices in turbulence evolve by interacting with each other, being deformed, elongated, reconnected, rotated, and translated by mutual interactions, sometimes consisting of vortex tangles.
Such an evolution of vortices is what we expect in the inviscid limit of classic fluid turbulence.
(See Saffman\cite{Saffman1992}, for example.)

The role of a quantum vortex or quantum turbulence in the heat transport of helium-4 is also an important issue in science and engineering\cite{Bao2019,Tatsumoto2002}. 
Although quantum vortices generated in a superfluid helium-4 are considered to suppress heat transport, the detailed mechanism of heat transport is not yet fully understood. 
To clarify the mechanism, identifying individual quantum vortices and understanding their dynamics are essential.

Quantum vortex dynamics have been studied extensively both by experiments\cite{Bewley2006,Paoletti2008,Paoletti2010,Chagovets2011,Mantia2013,Mantia2014a,Mantia2014b,Mantia2016,Kubo2017,Mastracci2018,Kubo2019,Tang2021,Svancara2021,Chen2022a,Chen2022b,Sakaki2022a,Sakaki2022b} and by numerical simulations\cite{Schwarz1985,Adachi2010,Boue2013,Mineda2013,Yui2015,Yui2020,Nore1997,Kobayashi2005,Kobayashi2006,Zuccher2012,Villois2016b,Stagg2017,Yoshida2019}.
Among these approaches, numerical simulations of the Biot-Savart model and the Gross-Pitaevskii (GP) equation can give us detailed information on quantum vortex dynamics.

The Biot-Savart model represents a quantum vortex as an ideal limit of a vortex filament in a classic fluid, and has been studied extensively\cite{Schwarz1985,Adachi2010,Boue2013,Mineda2013,Yui2015,Yui2020}.
However, this model ignores the vortex core, which is one of the characteristics in a quantum vortex and an important mechanism on a small scale.
Because of this nature, artificial operations appear inevitably in important events such as vortex reconnection and annihilation. 

The GP equation captures more of the detailed vortex dynamics, including the small scale events mentioned above, more naturally than the Biot-Savart model because the GP equation resolves a vortex-core scale which is characterized by the healing length $\xi$.
This element of the GP equation allows us to study vortex reconnection, creations and annihilation, energy spectrum, and energy transfer among scales (spectral dynamics) associated with elementary vortex dynamics, without adopting additional artificial models.

While GP simulations have been carried out extensively in these contexts\cite{Nore1997,Kobayashi2005,Kobayashi2006,Berloff2007,White2010,Zuccher2012,Villois2016,Villois2016b,Stagg2017,Yoshida2019,Yoshida2022,Zuccher2012,HussainDuraisamy2021}, studying statistical attributes of quantum vortices is another important subject which can be examined by a GP simulation.
Quantum vortex statistics, such as the vortex core radius, the distance between vortices, and the curvature of vortices, are closely related with experimental observation (see Ref.\cite{Sakaki2022a}, for example).
Thus clarifying the statistics through numerical simulations serves to explain the physics of quantum turbulence considerably.

In a GP equation, a quantum vortex can be extracted from the numerical data of a GP simulation by finding the zero-crossing of a numerical solution ${\psi} \in \mathbb{C}$.
In other words, a line of intersection of $\Re(\psi)=0$ and $\Im(\psi)=0$ surfaces is called a vortex axis.
However, the GP equation does not directly construct a quantum vortex axis.
Thus we consider developing an identification method of vortex axes and enable extracting vortex statistics out of the numerical data of GP simulations which are carried out to study the spectrum and/or heat transport of quantum turbulence.

In the last few decades, methods to identify three-dimensional (3D) quantum vortex axes in GP simulations have been developed\cite{Taylor2014,Proment2013,Krstulovic2012,Villois2016}.
Krstulovic\cite{Krstulovic2012} identified a vortex axis using an iterative method;
detecting the points of a vortex axis satisfying $\psi=0$, using the Newton-Raphson (NR) method. 
This was followed by a computation of the spectrum of the Kelvin waves on the vortex axes, and some other statistics associated with vortex axes.
Villois et al.\cite{Villois2016} found points on vortex axes using the  expansion of $|\psi|^2$ and the NR method, and connected the points by extending a vortex axis in the direction of the pseudo-vorticity vector, which can be expected to give an approximate direction tangential to a vortex axis.
Although these methods can give a good result in the identification of vortex axes,  a sequential repetition of an iterative NR method may not be suitable for the analysis of large-scale simulation data.
In addition, we consider that tool sets for for studying statistics of vortices should be developed further, so that we can extract turbulence statistics out of ensemble of individual quantum vortices in a GP simulation.

Hussain and Duraisamy \cite{HussainDuraisamy2021} also utilize a concept of a pseudo-vorticity which is used in Villois\cite{Villois2016}.
However, this work is limited to a couple of vortices.
Furthermore, while the pseudo-vorticity can represent the direction parallel to a vortex axis well, it does not necessarily give direct information of locations of vortex axes.
In this sense, a vortex identification method which represents geometrical properties of swirling vortices more directly can be helpful for a vortex identification of quantum turbulence, especially in a large-scale computation.

In this article, we develop a vortex identification method such that a position of a vortex axis can be obtained together with practically sufficient reliability, and that the method is applicable to large-scale numerical simulation data, by applying vortex identification methods used in classical turbulence\cite{Miura1997,Kida1998} with an appropriate modification for a quantum vortex.
We apply this method to extract quantum vortices out of GP simulation data, for the purpose of verifying the implementation, validating its physical appropriateness, characterizing vortices by various statistics, and elucidating the physics of quantum vortices.
It is emphasized here that applying a vortex identification method established in studies of classic fluid turbulence is meaningful in understanding elementary vortex dynamics common in quantum and classic fluid turbulence.
Since the methods in Refs.\cite{Miura1997,Kida1998} have been used in various turbulence studies\cite{Kida1998b,Makihara2002,Miura2002,Miura2004,Kawahara2005,Goto2009,Oka2021,Matsuura2022}, applying this method to quantum vortex dynamics is expected to enable studying vortex dynamics fit for this purpose.
In other words, we can study vortices in both quantum and classic turbulence from a unified point of view.

This article is organized as follows.
In section 2, the outline of our vortex identification method is presented, together with a modification from the original method for studying quantum vortices.
In section 3, quantum vortex axes are identified from GP simulation data. 
Some statistics obtained through the identification are reported in section 4.
Section 5 is the summary of this article.

\section{Identification of a quantum vortex}
The GP equation can be described as
\begin{equation}
\label{eq:GP}
i\hbar\frac{\partial}{\partial t}\psi
   =-\frac{\hbar^2}{2m}\nabla^2\psi-\mu\psi+g|\psi|^2\psi,
\end{equation}
where $m$ is the mass of the boson, $g$ is the coupling constant, $\mu$ is the chemical potential and $h$ is the Planck constant divided by $2 \pi$.
Equation (\ref{eq:GP}) is normalized in the same way as Refs.\cite{Yoshida2006,Yoshida2019,Yoshida2022}.
The coherent length $\xi$ satisfies $\xi=\hbar/\sqrt{2mg\bar{\rho}}$, where $\bar{\rho}$ is the space average of the $\rho=|\psi|^2$.

A quantum vortex axis in a solution ${\psi} \in \mathbb{C}$ of the GP equation is defined as a line of intersection of $\Re(\psi)=0$ and $\Im(\psi)=0$ surfaces\cite{Gross1961,Pitaevskii1961}.
In this study, the vortex identification method developed in classical turbulence\cite{Miura1997,Kida1998} is applied in order to detect such a line.

\begin{figure}[h]
\begin{center}
  \includegraphics[width=0.45\textwidth]{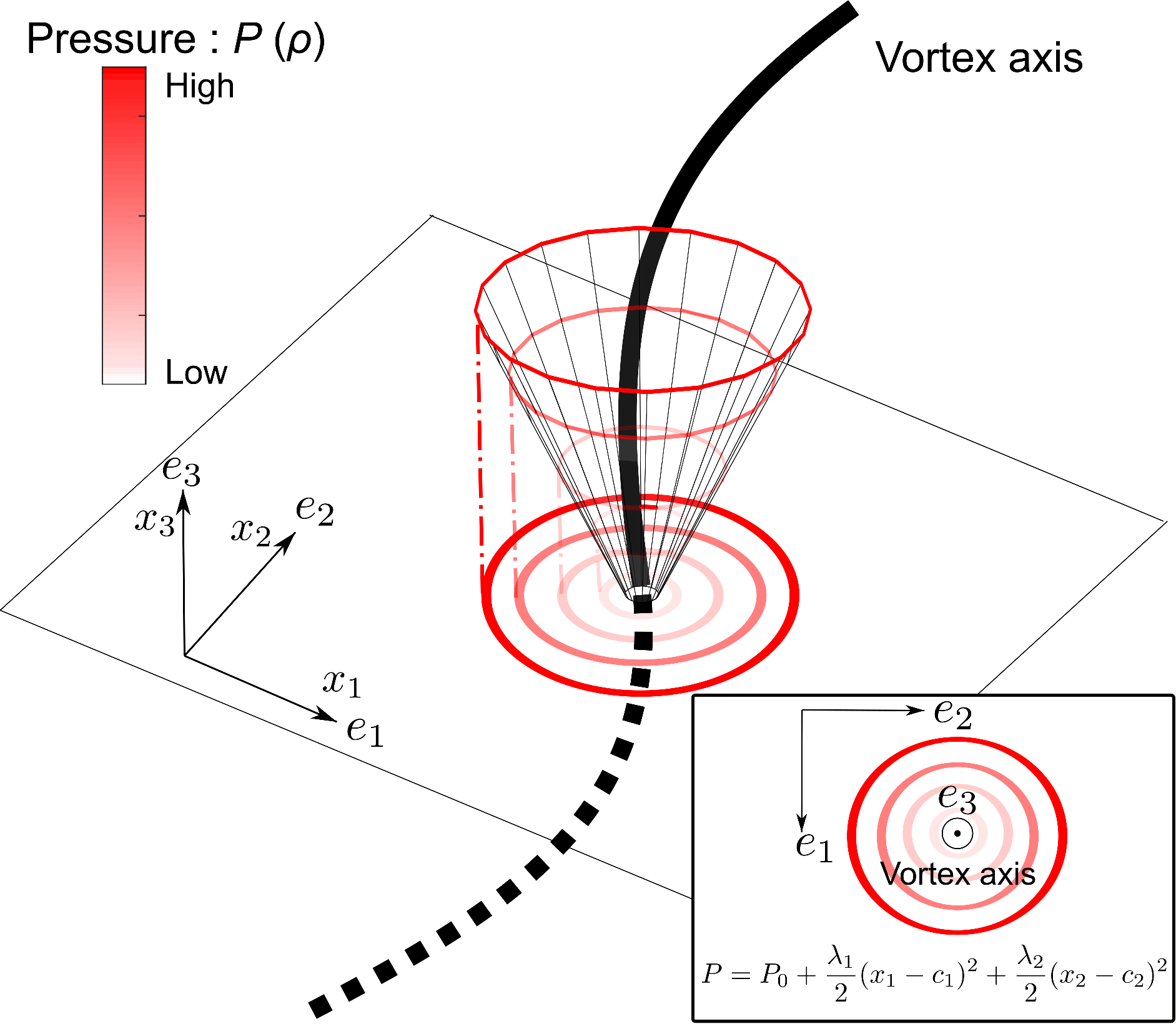}
  \caption{Schematic view of vortex identification by the use of the pressure Hessian.\cite{Miura1997}.
    Pressure $P$ is replaced by mass density $\rho$ in a case of a quantum vortex identification. 
In the case of a quantum vortex, an ellipse is almost a true circle.}
\label{fig:Miura1997}   
\end{center}
\end{figure}

The identification method  for classic turbulence in Refs.\cite{Miura1997,Kida1998} can be  summarized briefly as follows.
Figure \ref{fig:Miura1997} shows a schematic view of this method.
Firstly,  a candidate point of a vortex axis, a plane of swirling motions, and the direction tangential to a vortex axis are computed, based on eigenvalues and eigenvectors of the Hessian of the pressure $P$ at each grid point of a numerical simulation.
From the symmetry of the Hessian, the eigenvalues $\lambda_i\left(i=1,2,3\right)$ are real values and can be set $\lambda_1\geq\lambda_2\geq\lambda_3$ without loss of generality. 
By a local  coordinate rotation,  the pressure can be expressed approximately as
\begin{eqnarray} 
P \sim P_0+\frac{\lambda_i}{2}\left(x_i-c_i\right)^2 \label{eq:Press}
\end{eqnarray}
(we take the summation for $i=1$ to $3$ for a repeated surface), where $x_i$ are the coordinates in the direction of the three eigenvectors, $\bm{e}_i$ corresponding to the eigenvalue $\lambda_i \;(i=1,2,3)$, and the three eigenvectors are orthogonal to each other.
For $\lambda_1\geq\lambda_2>0 $, the pressure $P$ at $(c_1,c_2)$ is locally minimum on a $\left(x_1,x_2\right)$-plane (a plane of swirling motions), which is perpendicular to the $\bm{e}_3$ vector (the direction tangential to a vortex axis).
See Eqs.(2.3) and (2.4) in Ref.\cite{Miura1997} to determine $P_0$ and the position of vortex axes $(c_1,c_2,c_3)$.
Secondly, some points which satisfy the condition above, but are not followed by swirling flows, are excluded by the use of the swirling condition (a condition that the streamline relative to the point $(c_1,c_2,c_3)$ is circular).
Finally, by connecting the resultant set of the $(c_1,c_2,c_3)$ to the direction of the third eigenvalue of the Hessian, the vortex axes are obtained. 

This vortex identification method is applied to a GP simulation in three steps, as follows.
Firstly, by the definition of vortex axes $\rho=0$, we can expect  $|\psi|^2 < \epsilon^2$ ($\epsilon$ is an appropriately small positive number)  on the grid points in the neighbourhood of the vortex axes.
By the use of this condition  $|\psi|^2 < \epsilon^2$,  we can focus on a relatively small number of grid points to find candidate points of the vortex axes.
(A detailed value of $\epsilon$ depends fully on the normalization of $\psi$ and excitation of the Fourier coefficients of $\psi$, in a high-wavenumber region.)

Secondly, we follow the method for classic turbulence introduced above.
The pressure $P$ and the local-minimum value of the pressure $P_0$ are replaced by the mass density $\rho=|\psi|^2$ and $\rho_0$, respectively.
For expecting $\rho(c_1,c_2,c_3)=0$ on a point on a vortex axis $(c_1,c_2,c_3)$, we demand $P_0 = \rho_0 \sim 0$ and  $\lambda_3 \sim 0$ within an order of the truncation error in Eq.(\ref{eq:Press}).
This makes a sharp difference from a classic incompressible fluid in which $\lambda_1 + \lambda_2 + \lambda_3 = 0$ without requiring $\lambda_3 \simeq 0$.
Here, we omit an iterative procedure in finding points of $\psi=0$, accepting the truncation error in Eq.(\ref{eq:Press}).
This shortcut of computation allows us to process many vortex axes in parallel easily, as we have done in classic fluid turbulence\cite{Miura2002,Miura2004}. 

Thirdly, we omit some erroneous points that satisfy the procedure above while they are not followed by swirling flows.
For this purpose, we need to impose a swirling condition.
Since a flow of $v_{\theta}=1/2\pi r$ with a singular pole at $r=0$ generated by a quantum vortex cannot be distinguished from other flow profiles correctly by the criterion in Ref.\cite{Kida1998}, or some other well-established criteria using the velocity gradient tensor such as so-called Q-criterion, we introduce a new, simplified swirling condition.
Since the velocity is almost a true circle if $(c_1,c_2)$ is on a vortex axis and $r$ is set reasonably, we consider a circle of radius $r$ at the center $(c_1,c_2)$ on the $\bm{e}_1-\bm{e}_2$ plane, and check the sign of the inner product $\bm{v}\cdot \bm{e}_{\Theta_j}$ at some points.
We consider that $(c_1,c_2)$ is a point of a vortex axis when $\bm{v}\cdot \bm{e}_{\Theta_j}$ takes the same sign along the circle.

Finally, by connecting those candidate points that satisfy the swirling conditions in the $\bm{e}_3$ direction, we obtain the vortex axes.
The candidates for a vortex axis satisfy the requirement for a quantum vortex $\Re\left(\psi\right)=\Im\left(\psi\right)=0$ within a range of the truncation error in Eq.(\ref{eq:Press}).

Many of the procedures such as computations of the mass-density Hessian eigenvalues and eigenvectors, and candidates for vortex points $\left(c_1,c_2,c_3\right)$ are suitable for massive parallel computations. 
This is a big advantage of our method compared to other similar vortex identification methods of GP turbulence which use a recursive process to extract points of $\Re(\psi)=\Im(\psi)=0$ from simulation data, because the points to construct vortex axes can become very great in large-scale numerical simulations.

We emphasize that the procedure for a GP simulation presented here is quite similar to that for classic turbulence simulation\cite{Miura1997,Kida1998}.
This enables us to compare various aspects of vortices in a GP simulation to those in classic turbulence from a unified point of view.
While some procedures in our identification method share those in Ref.\cite {Villois2016}, our method presents a clearer geometrical character of vortices as a local basin of low or zero mass density of a quantum fluid, corresponding to that of the pressure of a classic fluid.

\section{Analysis of vortices in a GP simulation}
\label{sec:3}
\subsection{GP simulation}
In this subsection, the outline of a GP simulation is presented.
Equation (\ref{eq:GP}) is solved numerically under the triple periodic boundary condition in the domain $\left[0,2\pi\right)^3$ by the pseudo-spectral method and the Runge-Kutta-Gill scheme.
The number of grid points is ${256}^3$, and thus the grid width is $2\pi/256\approx2.45\times{10}^{-2}$.
The time step $\Delta t$ is $5.0\times10^{-4}$.
The coherent length $\xi$ is $6.25\times{10}^{-2}$.
We impose an external dissipation in a large-wavenumber range as $D_k=-\nu k^4$, following the method in Refs.\cite{Yoshida2019,Yoshida2022}.
The parameter $\nu$ is $2.5\times10^{-4}$.
See Refs.\cite{Yoshida2019,Yoshida2022} for details of the numerical simulation and earlier results of our developed GP turbulence simulations.
We note that a very precise spectral computation is not required in our vortex identification.
We use the second-order finite differential scheme for a computation of the Hessian of $\rho=|\psi|^2$, and omit a spectral computation there.

\begin{figure}[h]
\begin{center}
  \includegraphics[width=0.3\textwidth]{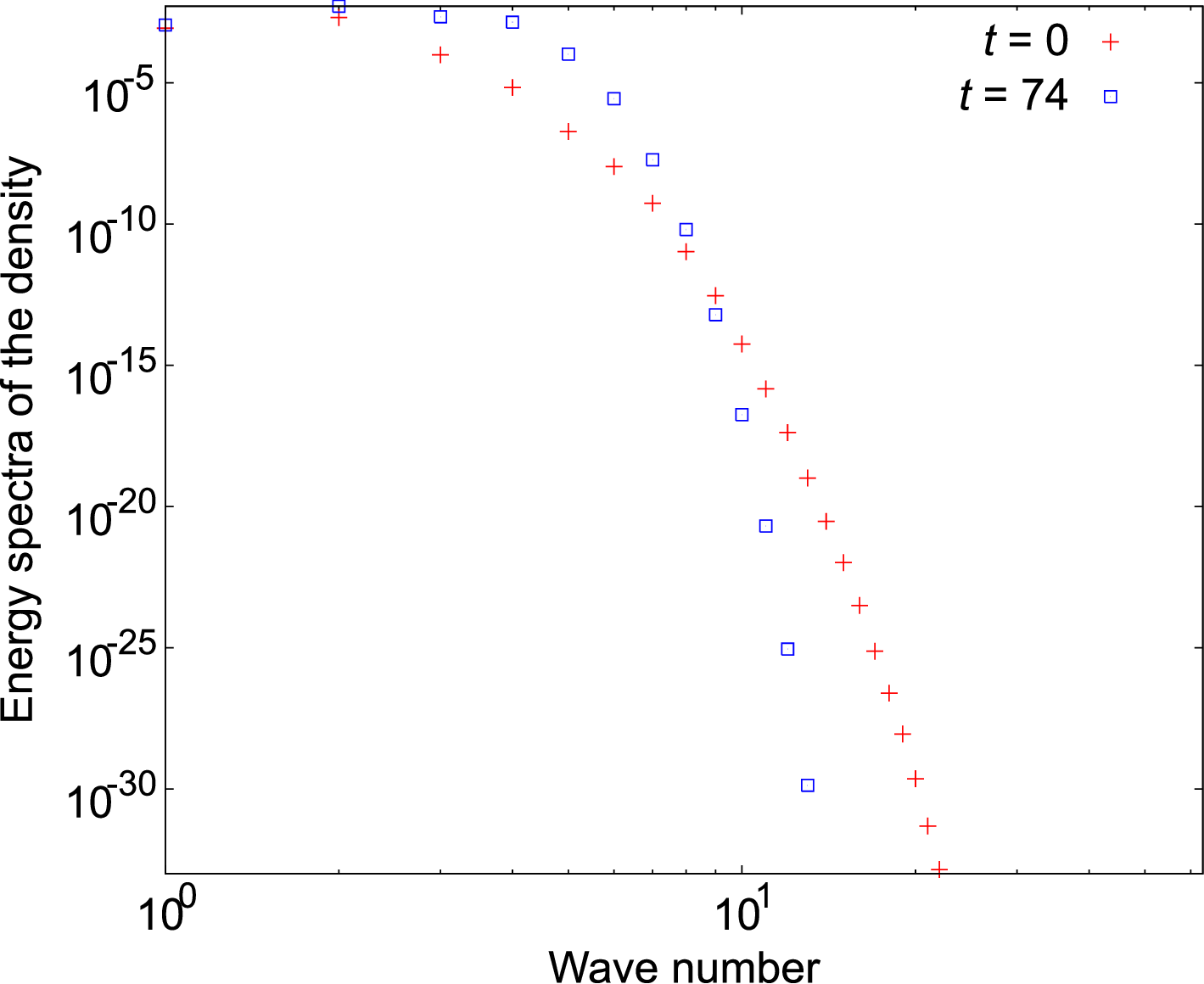}
\caption{The time development of the spectra of the density. The red cross and blue square are the spectra of the initial $(t =\ 0)$ and final time $(t= 74)$, respectively.}
\label{fig:spec}   
\end{center}
\end{figure}

In Fig.\ref{fig:spec}, the spectrum of the mass density $\rho=|\psi|^2$ is presented at the initial (red cross) and the final (blue square) time.
By introducing $D_k$ at the right-hand side of Eq.(\ref{eq:GP}), a high-wavenumber regime of the mass density is suppressed. 
This makes $\psi$ smooth and provides an appropriate test ground for the initial assessment of our identification method for vortex axes.
Although turbulence is not fully developed in this simulation, the basic events of quantum vortices such as vortex translation, reconnection, and collapse of a vortex ring are contained in the present simulation data.
(We exclude figures of the events for simplicity.)
%

%
%
\begin{figure}[h]
    \begin{tabular}{c}
      \begin{minipage}[t]{\hsize}
        \centering
        \includegraphics[width=0.5\textwidth]{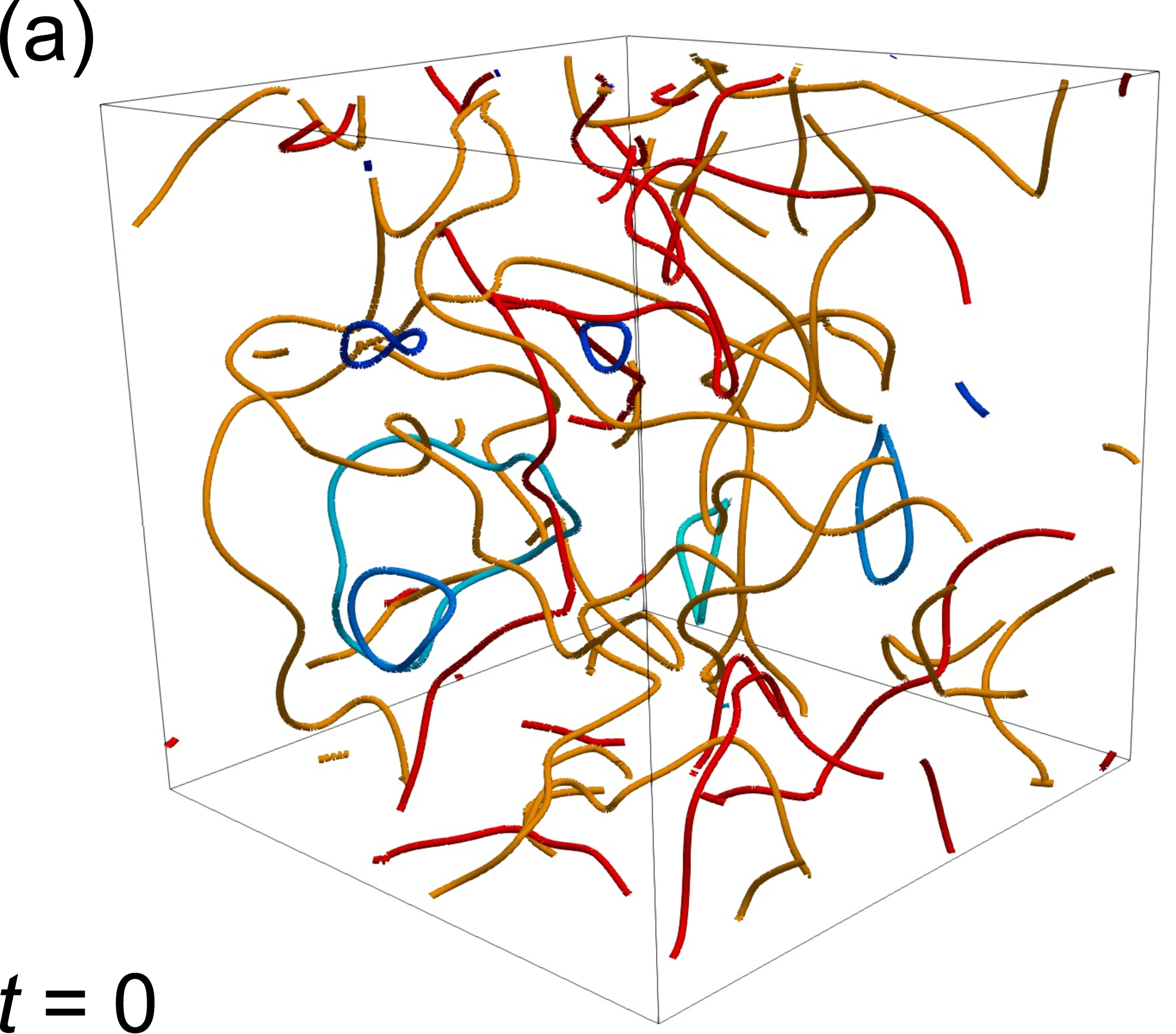}
      \end{minipage} \\
      \begin{minipage}[t]{\hsize}
        \centering
        \includegraphics[width=0.5\textwidth]{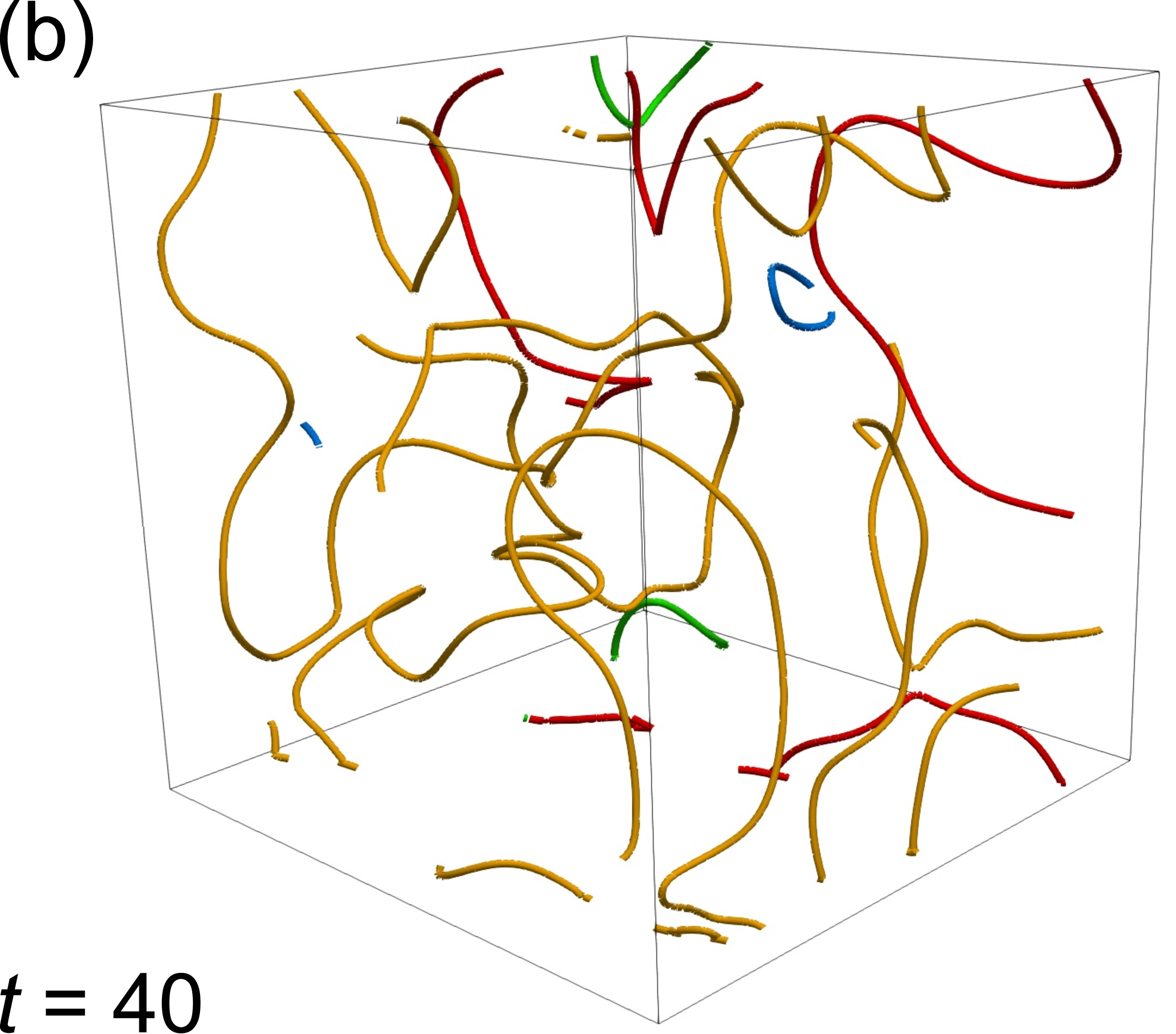}
      \end{minipage} \\
      \begin{minipage}[t]{\hsize}
        \centering
        \includegraphics[width=0.5\textwidth]{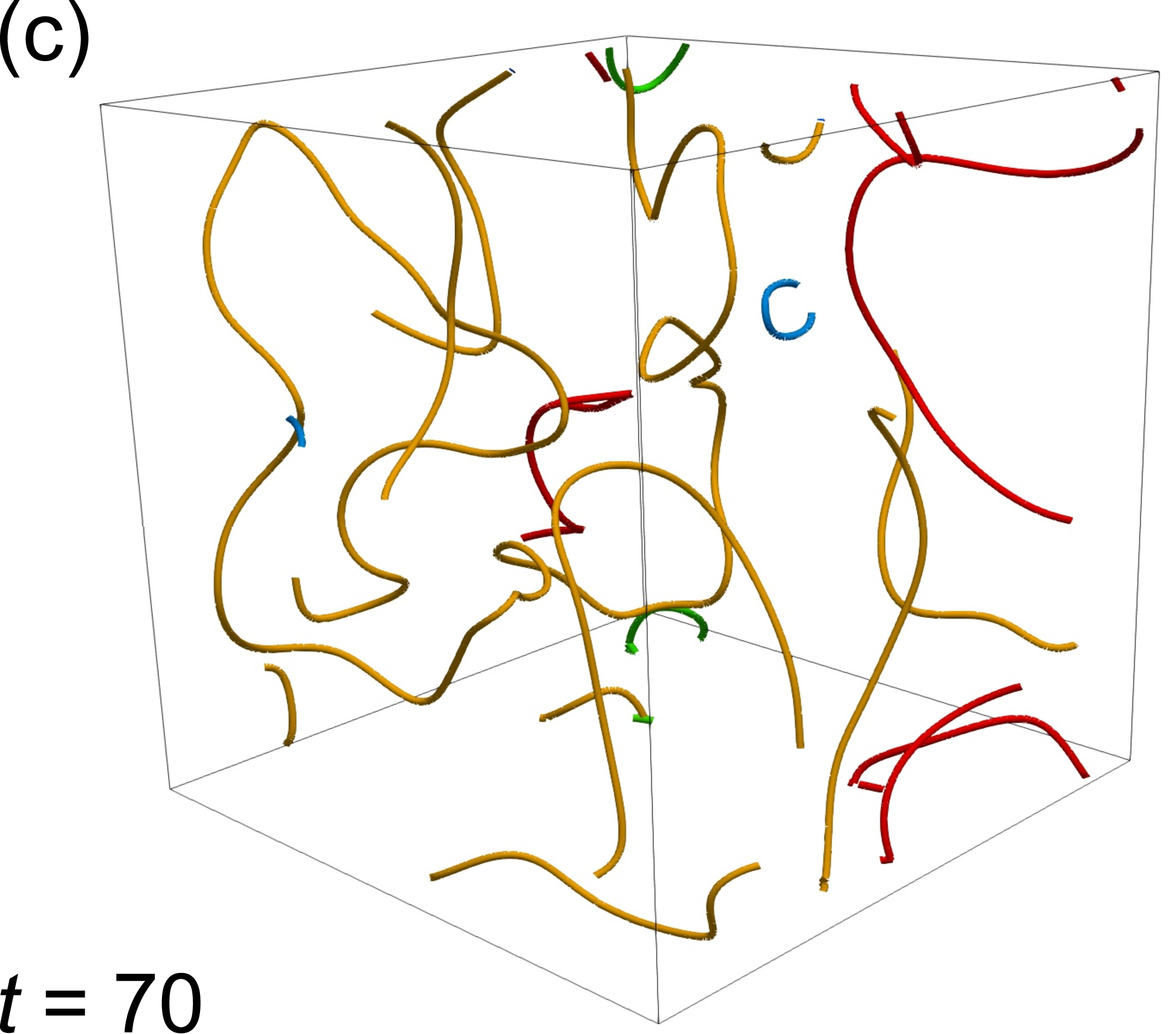}
      \end{minipage} 
    \end{tabular}
    \caption{The vortices are identified by the present method. In each time step, a different color represents a different vortex. Time in (a) is 0, (b) is 40, and (c) is 70.}
    \label{fig:visu}
  \end{figure}

\subsection{Identification of vortex axes in a turbulence simulation}

\begin{figure}[h]
\begin{center}
  \includegraphics[width=0.3\textwidth]{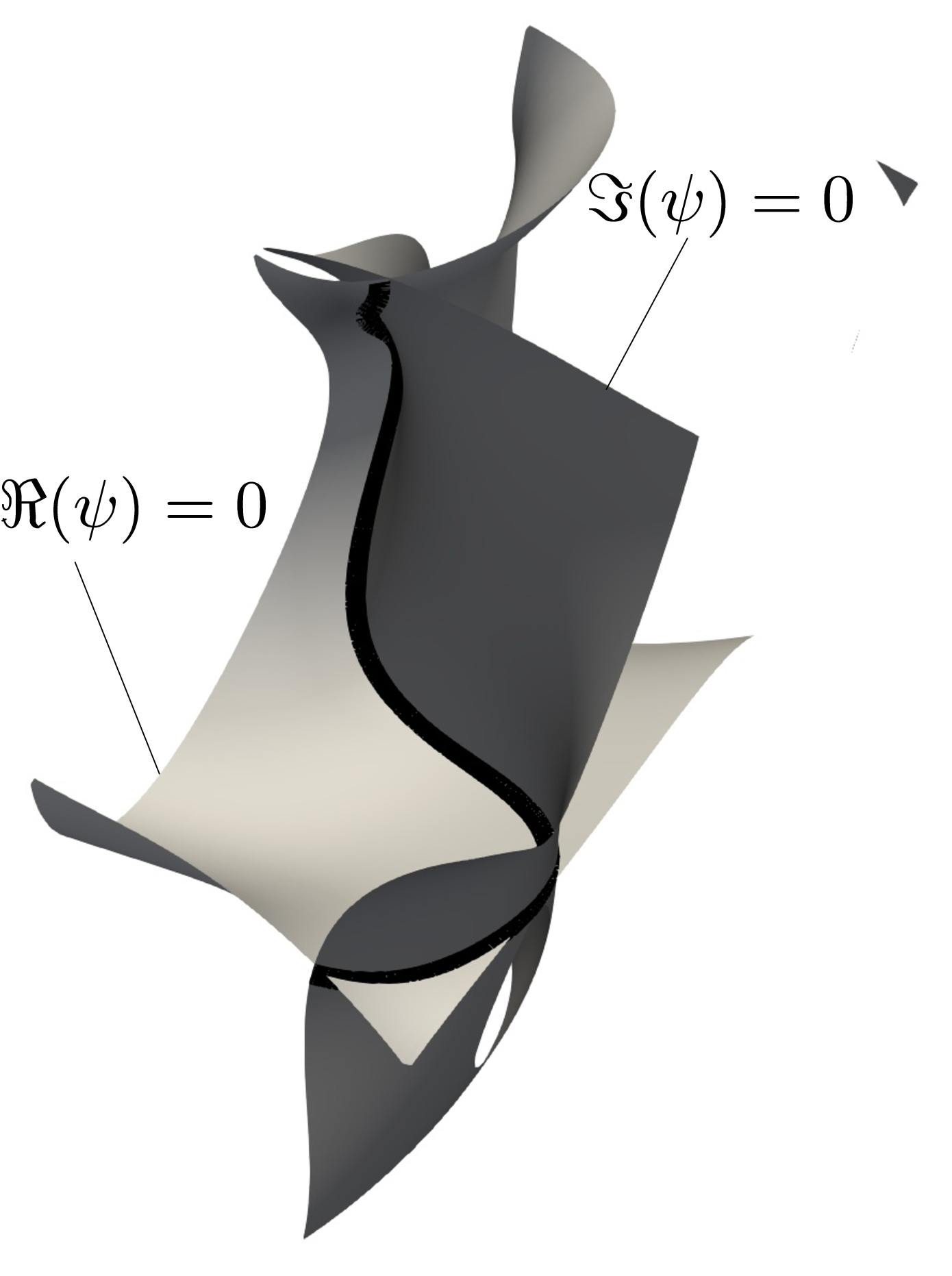}
  \caption{The vortex is the same time as that in Fig.\ref{fig:visu} (c) in the $0\le x\le\pi/2$, $\pi\le y\le2\pi$, and $0\le z\le\pi$. The thick black line is a vortex axis.
    The light and dark grey surfaces are $\Re\left(\psi\right)=0$ and $\Im\left(\psi\right)=0$, respectively.}
\label{fig:magnify}   
\end{center}
\end{figure}
\begin{figure}[h]
\begin{center}
  \includegraphics[width=0.5\textwidth]{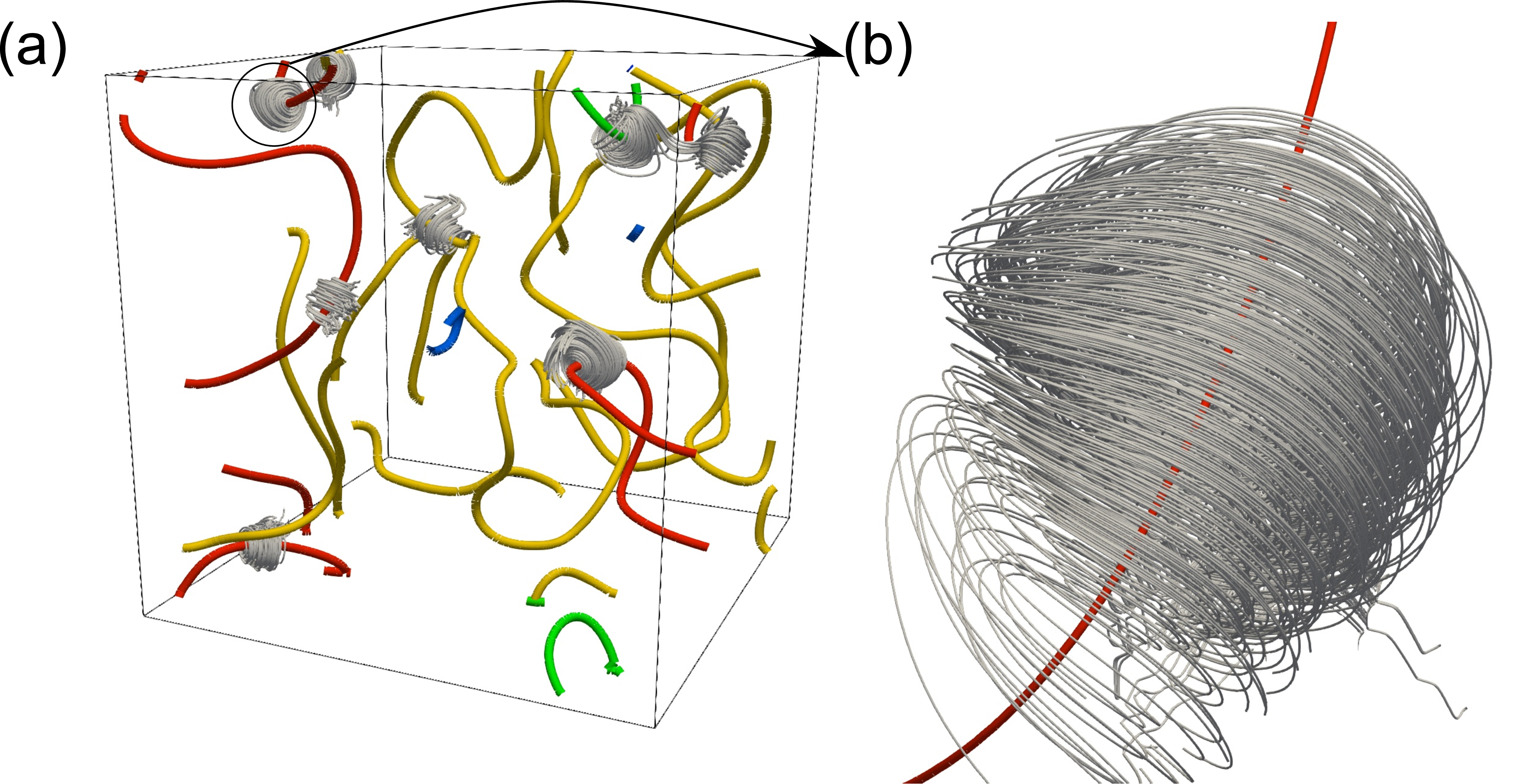}
  \caption{In (a), the swirling flow (light grey) around the vortex axis is visualized at the time in Fig.\ref{fig:visu} (c). The color of the vortex axes is the same as Fig.\ref{fig:visu} (c).
    In (b), the streamline in the black circle of (a) is magnified.}
\label{fig:swirl_visu}   
\end{center}
\end{figure}

Figure \ref{fig:visu} shows the time development of vortex axes identified using our method. 
The color of a vortex axis in this figure  is set to distinguish one axis from the others, without any physical meaning.
Although many vortices appear tangled with each other, there are only nine in (a), four in (b), and four in (c) when the periodic boundary condition is considered in their identification.
The number of vortices is decreased by vortex ring collapses.

Figure \ref{fig:magnify} shows a magnification of the vortex axes (thick black line)  in Fig.\ref{fig:visu} (c) $(t=70)$ for the range of $0\le x\le\pi/2$, $\pi\le y\le2\pi$, and $0\le z\le\pi$ represented together with isosurfaces of $\Re\left(\psi\right)=0$ (light dark surfaces)  and $\Im\left(\psi\right)=0$ (dark grey surfaces). 
 It is confirmed in this figure that the quantum vortex axis exists on the line of $\Re\left(\psi\right)=\Im\left(\psi\right)=0$.
 
We verify here that swirling flows are actually formed around the vortex axes.
Figure \ref{fig:swirl_visu} represents streamlines (light grey) around the quantum vortices in Fig.\ref{fig:visu} (c) $(t= 70)$. 
The color of the vortex axes is the same as in Fig.\ref{fig:visu} (c). 
This figure indicates that there are swirling motions around the vortex axes identified by our method, and indicate that these vortex axes are what we mean by quantum vortices.
\begin{figure}[h]
\begin{center}
  \includegraphics[width=0.3\textwidth]{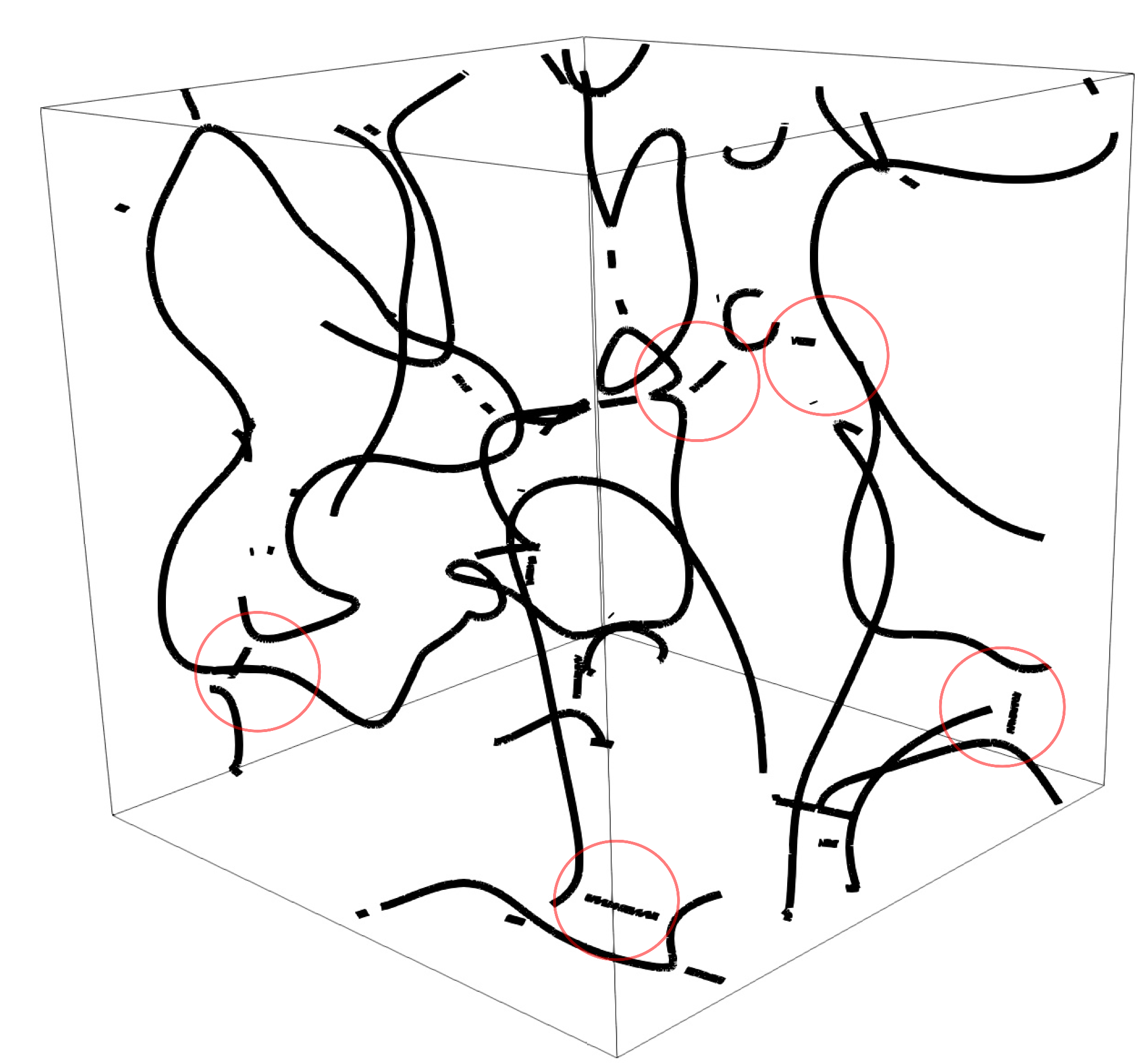}
\caption{
The vortices are identified without the swirling condition in Fig.\ref{fig:visu} (c) $(t = 70)$. 
The vortex axes are represented by the black line. In the red circle, there is no swirling flow around them. 
However, in this area, $\Re \left(\psi\right)^2$ and $\Im \left(\psi\right)^2$ are much smaller and the condition of the Hessian of the density is satisfied.}
\label{fig:pseudo_vortex}   
\end{center}
\end{figure}

It is worth examining the importance of the swirl condition here.
Figure \ref{fig:pseudo_vortex} shows vortex axes identified without the swirling condition by the same data set as that used in Fig.\ref{fig:visu} (c) $(t=70)$. 
In Fig.\ref{fig:pseudo_vortex}, the axes surrounded by the red circles do not satisfy the swirling condition.
We have confirmed that there is no swirling flow around them, although both a filtering condition $\rho=|\psi|^2 < \epsilon^2$  and the condition of the Hessian of the mass density $\lambda_1\geq\lambda_2>0$ are satisfied.
The existence of these encircled points indicates that the two conditions are insufficient to identify the vortex points.
These encircled points satisfy $|\psi|^2 < \epsilon^2$ and $\lambda_2 > 0$.
This means that $\psi \sim 0$ within the range of the truncation error of Eq.(\ref{eq:Press}) on these  points.
Since eliminating these points by changing $\epsilon$ or by the Taylor expansion of $\rho$, our new swirling condition plays a crucial role in the elimination.

\section{Vortex statistics}
\label{sec:4}
Now we move on to the analysis of vortex statistics obtained from a GP simulation using our vortex identification method.
%

\subsection{Vortex length, distance, and radius}
\begin{figure}[h]
\begin{center}
  \includegraphics[width=0.3\textwidth]{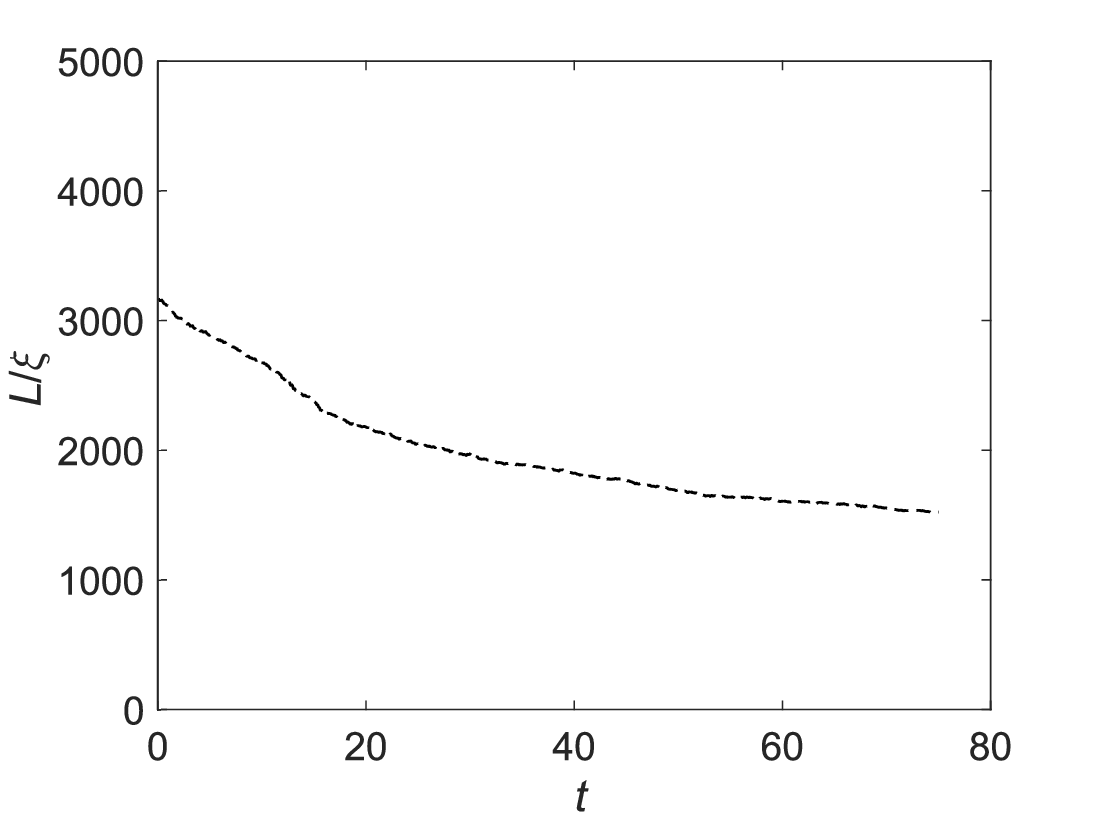}
\caption{The time development of the total vortex length $L$ is represented by the black dotted line. The scale is normalized by the coherent length $\xi$. $L$ decreases with time.}
\label{fig:vortex_length}   
\end{center}
\end{figure}
\begin{figure}[h]
\begin{center}
  \includegraphics[width=0.3\textwidth]{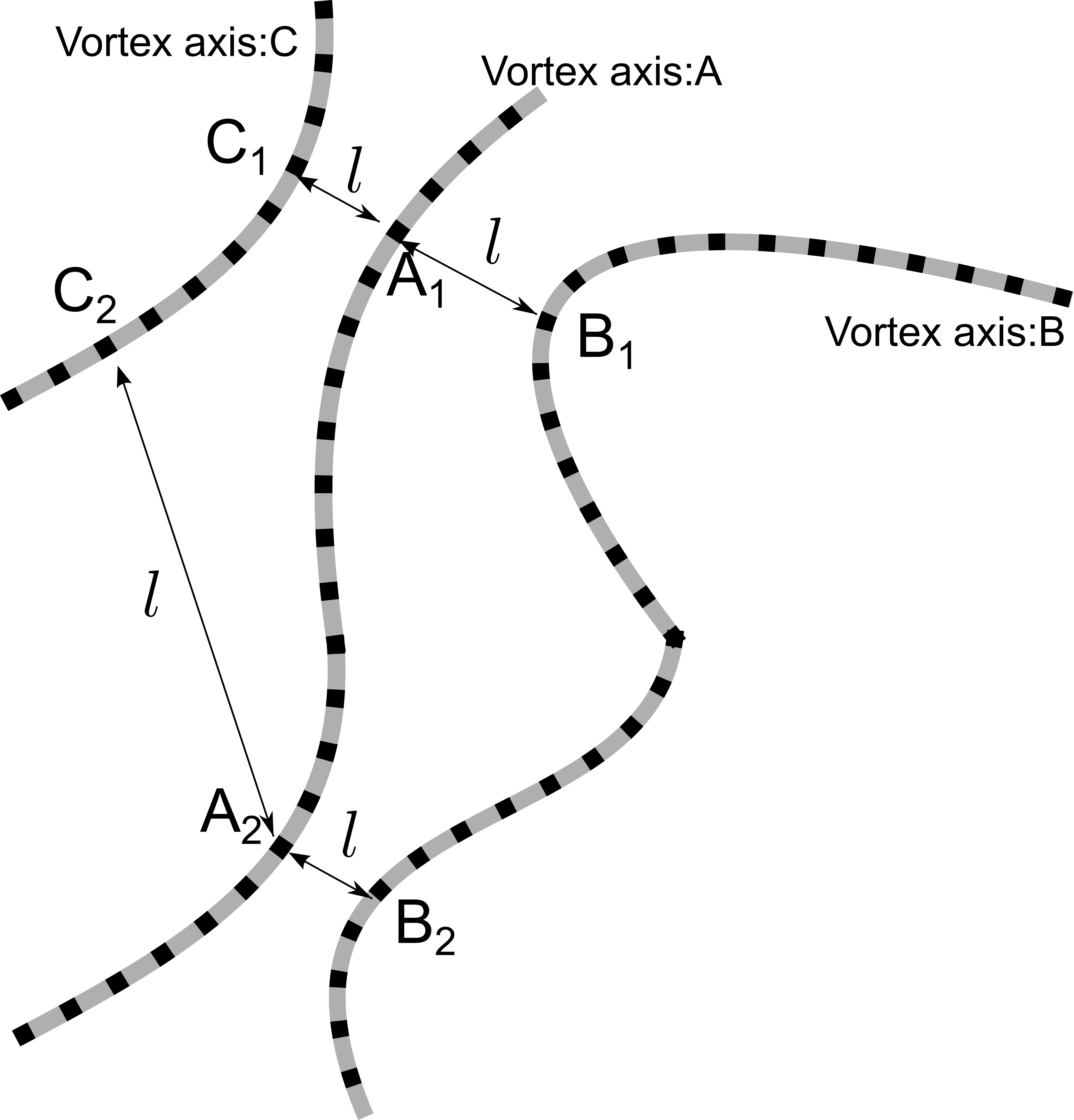}
  \caption{The schematic view of the vortex distance $l$. The vortex distance $l$ is defined for each vortex point as the minimum distance from one vortex point to another  that is not the element of the same vortex axis.}
\label{fig:def_min_length}   
\end{center}
\end{figure}
\begin{figure}[h]
\begin{center}
  \includegraphics[width=0.3\textwidth]{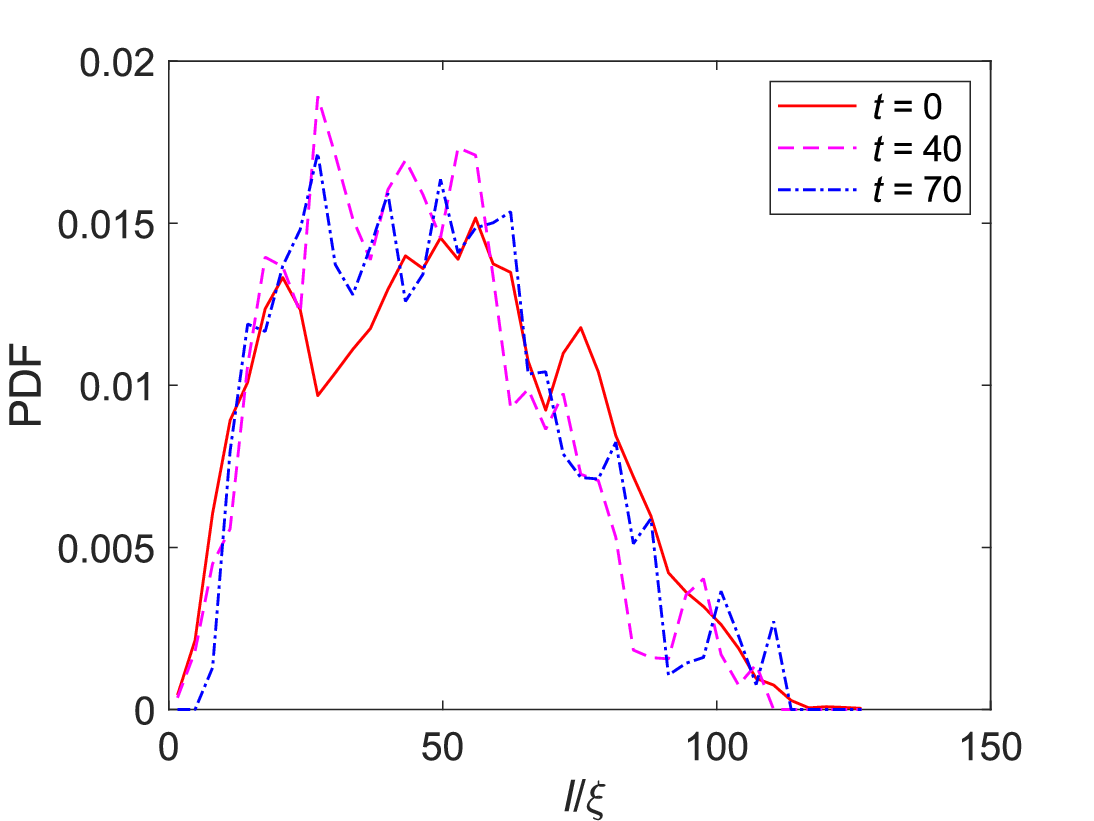}
\caption{The probability density functions of the vortex distance in $t=0$, $t=40$, and $t=70$. The scale is normalized by the coherent length $\xi$.}
\label{fig:PDF_length}   
\end{center}
\end{figure}
\begin{figure}[h]
\begin{center}
  \includegraphics[width=0.3\textwidth]{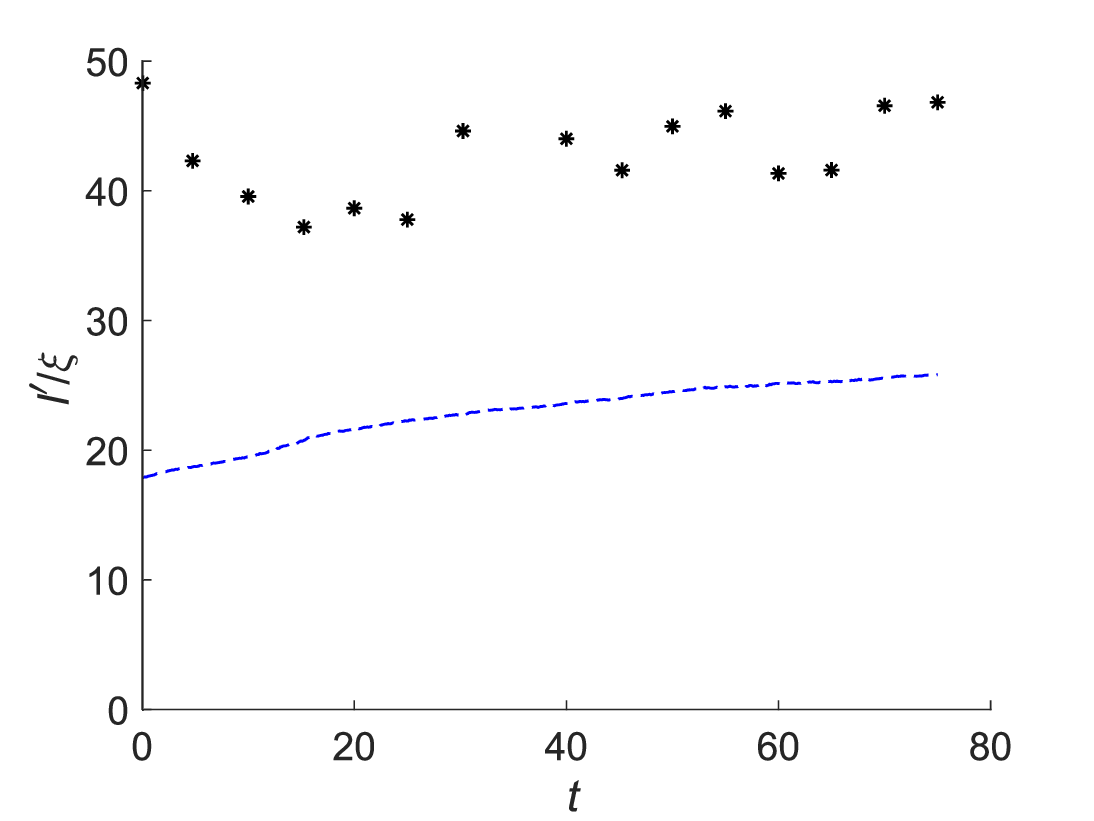}
\caption{The time development of the distance of vortices estimated by dimensional analysis $l^{\prime}$ is represented by the dotted blue line. 
The scale is normalized by the coherent length $\xi$. 
The black asterisk is the expected value $\bar{l}$ calculated by the PDFs in Fig.\ref{fig:PDF_length}.}
\label{fig:compare_PDF_dim}   
\end{center}
\end{figure}
\begin{figure}[h]
\begin{center}
  \includegraphics[width=0.3\textwidth]{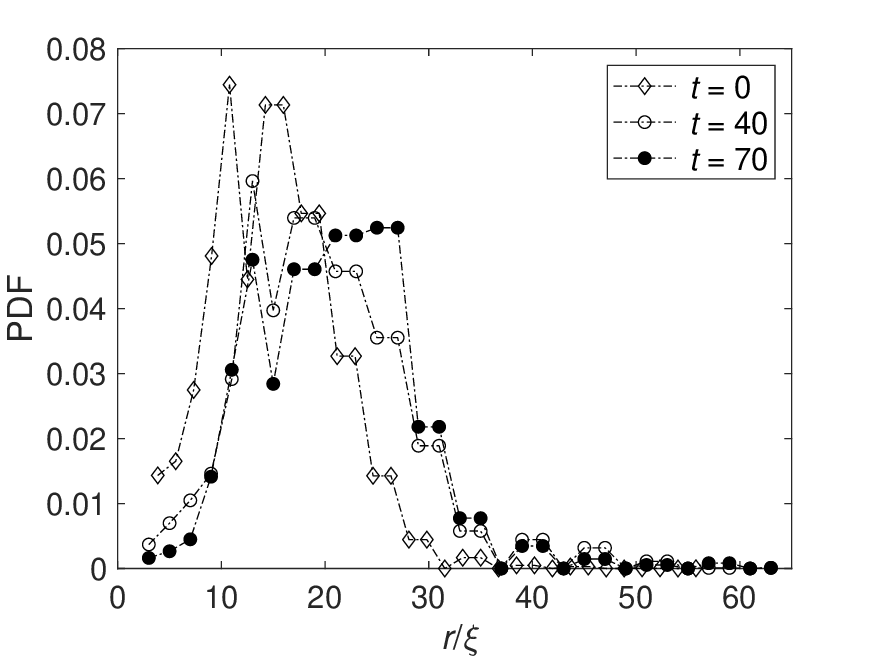}
\caption{The PDFs of the radius of a swirling region of vortices are represented by a black unfilled diamond (at $t=0$), black unfilled circle (at $t=40$), and black filled circle (at $t=70$).
The scale is normalized by the coherent length $\xi$. }
\label{fig:swirl_rad}   
\end{center}
\end{figure}

Figure \ref{fig:vortex_length} shows the time development of the total vortex length $L$, normalized by the coherent length $\xi$.
The normalized length $L/\xi$ decreases with time, as $\rho=|\psi|^2$ decays monotonically, and slowly especially for $t \geq 40$. 
Combining the observations of the vortex axes in Fig.\ref{fig:visu} and Fig.\ref{fig:vortex_length}, the main transient phenomena such as a vortex reconnection and a collapse of the vortex ring are considered almost completed until $t\approx40$ and this simulation moves to a slowly-decaying, statistically quasi-steady state for $t \geq 40$.

%
The distance between two vortices also gives important information because a primary influence on a vortex motion comes from another, nearest-neighbouring vortex.
The vortex distance $l$ is defined for each vortex point as the minimum distance from one  vortex point to another that is not the element of the same vortex axis.
Figure \ref{fig:def_min_length} is a schematic view of the definition of $l$.

The probability density functions (PDF) of $l$ at $t =0$, $40$, and $70$ are shown in Fig.\ref{fig:PDF_length}.
The shape of the PDF is quite similar for the three snapshots.
The PDF for $t=40$ and $t=70$ in Fig.\ref{fig:PDF_length} have almost the same shape.
Using the PDF $P(l)$ at each time of evolution, the mean vortex distance (the shortest distance from a point of vortex axis to the nearest-neighboring vortex) can be computed as 
\begin{equation}
\bar{l}=\int_{0}^{\infty}lP\left(l\right)dl \label{eq:PDF_expect}.
\end{equation}
We can also estimate the distance between the vortices $l^{\prime}$ from the vortex axes density $L/\left(2\pi\right)^3$ by a dimensional analysis as
\begin{equation}
l^{\prime}=\left(L/(2\pi)^3\right)^{-1/2}. \label{eq:dim}
\end{equation}
This estimation of the mean vortex distance $l^{\prime}$ is similar to that in an experimental study.
(See Ref.\cite{Mantia2014b}, for example.)

In Fig.\ref{fig:compare_PDF_dim}, the time evolution of $\bar{l}$ computed from the PDF of $l/\xi$ by Eq.(\ref{eq:PDF_expect}) and $l^{\prime}$ estimated by Eq.(\ref{eq:dim}) are represented by black asterisks ($*$) and a blue dashed line, respectively. 
The vortex distance $l^{\prime}$ obtained by the dimensional analysis (\ref{eq:dim}) is about half of the expected value $\bar{l}$ computed by Eq.(\ref{eq:PDF_expect}). 
This suggests that the nearest neighbour distance can be estimated as the half of the dimensional analysis in experiments.

In Fig.\ref{fig:swirl_rad}, the PDF of the radius of a swirling region of vortices is shown by a black unfilled diamond (at $t=0$), black unfilled circle (at $t=40$) and  (at $t=70$).
The scale is normalized by the coherent length $\xi$. 
The swirling region is defined here as the maximum extent of a swirling region around a vortex axis where the swirling condition in our vortex identification method is satisfied.
This swirling region corresponds to the so-called vortex core in the context of a classic fluid turbulence study\cite{Kida1998}. 
In fact, the PDF profile in Fig.\ref{fig:swirl_rad} is similar to that in Fig.3 of Ref.\cite{Kida1998}.

Figure \ref{fig:swirl_rad} also shows that the swirling region can be as large as $60 \xi$ as  maximum ($10\xi \sim 30\xi$ frequent sizes) at $t=70$.
This means that a swirling region can be much wider than the healing length $\xi$.
Since the pressure in the swirling region is lower than the surrounding region, a quantum vortex can capture a particle as large as $60 \xi$.
Although this numerical simulation is completed at $t=70$, the mean vortex extent can be larger if we continue this simulation for a sufficiently long time,
although the maximum extent $\sim 60 \xi$ can be limited due to the periodicity of the system (the system size is $2 \pi \sim 100\xi$).
We need to carry out a simulation with a small $\xi$ or a larger computational box, equivalently, to study the particle-capturing nature of vortices in GP turbulence.

Here we remember recent experimental reports on capture of an inertia particle by a quantum vortex\cite{Shukla2023}. 
On the one hand, it is reported that a single quantum vortex can capture an inertia particle (a marker particle used in a quantum vortex experiment), of which radius is as large as $10^5 \xi$.
On the other hand, there is another opinion that particle capture can be attributed to vortex bundles\cite{Sasa2011,Baggaley2012}.
The formation of a vortex bundle raises the possibility that a marker particle can be captured not by a single vortex but by a cluster of vortices that form the bundle.
While we do not have a concrete assertion on this subject yet, we emphasize that the observation in Fig.\ref{fig:swirl_rad} is meaningful because the maximum extent of the swirling region shrinks a huge gap between numerical simulations and experiments.
 
\subsection{Curvature of vortex axes} 
\begin{figure}[h]
\begin{center}
  \includegraphics[width=0.4\textwidth]{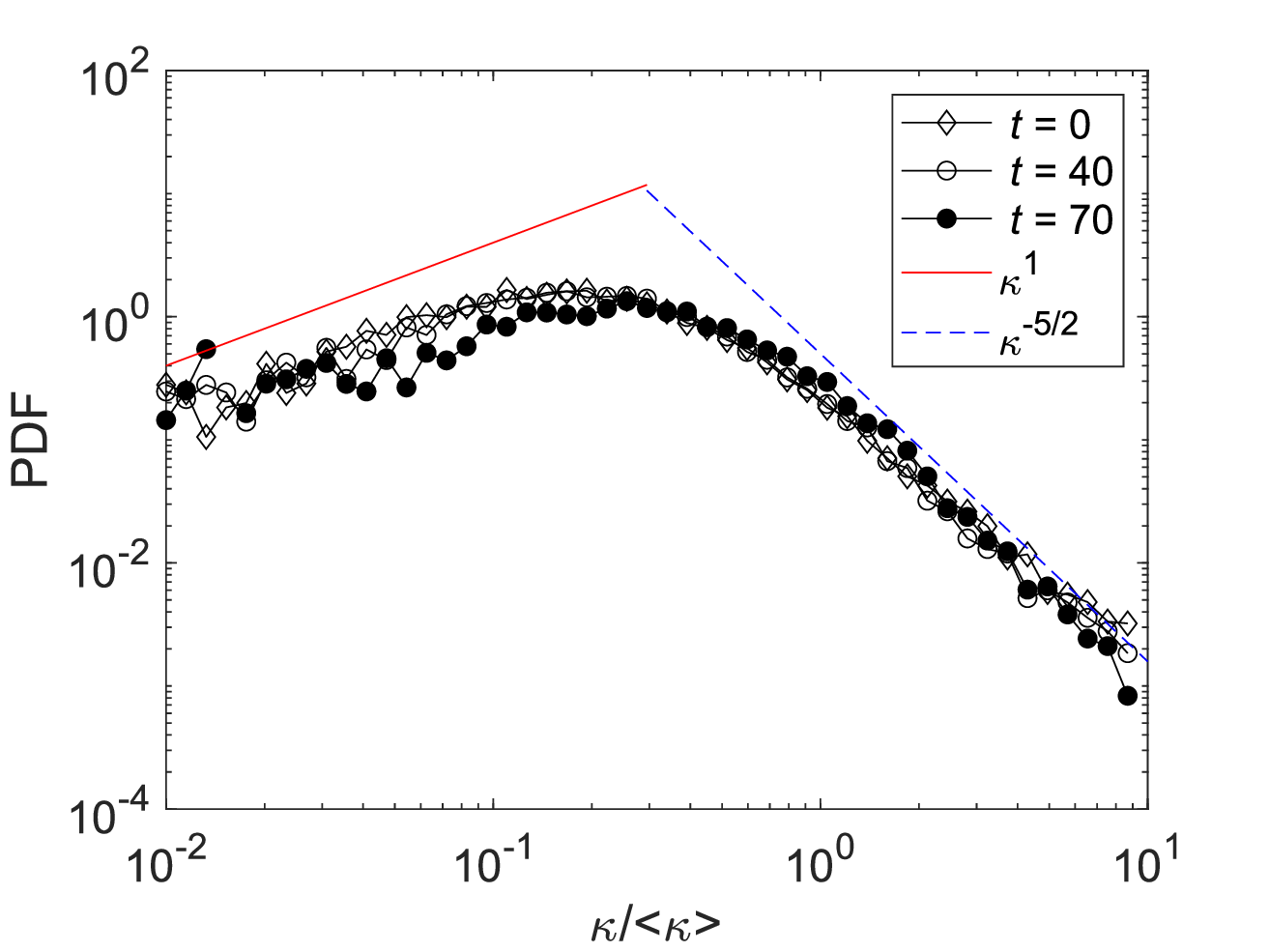}
\caption{The probability density function of the curvature $\kappa$ of vortex axes at $t=0$ (black unfilled diamond), $t=40$ (black unfilled circle), and $t=70$ (black filled circle). 
  The scale is normalized by the mean curvature $\langle\kappa\rangle$.
The red and blue dotted lines are the power-laws $\kappa^1$ and $\kappa^{-5/2}$, respectively.}
\label{fig:cuvature_3d}   
\end{center}
\end{figure}

As the final step of studying vortex statistics, we investigate the geometrical nature of vortex axes. 
Figure \ref{fig:cuvature_3d} is the PDF of the curvature $\kappa$ of vortex axes at $t=0$ (black unfilled diamond), $t=40$ (black unfilled circle), and $t=70$ (black filled circle).
The curvature is computed in the same way as the Biot-Savart simulations\cite{Yui2015,Yui2020}.
The PDF of the curvature has two power-laws, $\kappa^1$ in the small curvature region and $\kappa^{-5/2}$ in the large curvature region, regardless of the time.
The red and blue dotted lines are the power-law $\kappa^1$ and $\kappa^{-5/2}$, respectively.

The power-law $\kappa^1$ in the small curvature region is the same result as in Ref.\cite{Villois2016b} where the GP simulation includes developed and decaying turbulence.
In the literature\cite{Villois2016b}, the Taylor-Green flow was used as the initial condition, which is totally different from our simulation.
Since our simulation and that in Ref.\cite{Villois2016b} are different to each other in some important aspects, such as initial condition and forcing in the time evolution, a resemblance of the power-law $\kappa^1$ to that in Ref.\cite{Villois2016b} indicates that a shape of a curvature PDF in the small curvature region is independent of details of a numerical condition, whether the turbulence is developed or not.

On the other hand, the profile of the PDF at the large curvature region is different between our result and that in Ref.\cite{Villois2016b}.
This may be attributed to a strong suppression of high-wavenumber Fourier coefficients by hyperviscosity.
We need further investigation on this point by a larger numerical simulation.
Interestingly, however, the PDF profile at the large curvature region in our simulation coincides with that obtained in our recent experiments\cite{Sakaki2022a,Sakaki2022b}.
This suggests that the $\kappa^{-5/2}$ in our simulation represents some aspects of elongated quantum vortices as Refs.\cite{Sakaki2022a,Sakaki2022b}.
Since a diagnostic resolution is also limited in comparison to a real quantum vortex size of Angstrom, the appearance of $\kappa^{-5/2}$ may be attributed to a coarse-grained observations both in simulations and experiments.

\section{Concluding remarks}
We have developed a method to identify a quantum vortex in a GP turbulence simulation.
This method consists of a geometric condition based on eigenvalues and eigenvectors of the mass density Hessian, originally developed for classical turbulence in Refs.\cite{Miura1997,Kida1998}, and a new swirling condition that is applicable to a quantum vortex.
Our method successfully extracts vortex axes out of a GP simulation without using an iterative calculation, which often becomes a bottleneck in large-scale data analysis.
This method provides us with a unified approach to studying vortex statistics in the turbulence of both classic and quantum fluids.

We have also studied some statistics of vortices using the vortex identification method, in order to characterize  a quantum vortex in a GP simulation and elucidate some important physics.
The total length, mean distance, and maximum extent of swirling motions have been studied.
These characteristic lengths are related directly or indirectly to experimental observations and estimations of quantum vortices in experiments.
We have found that the maximum extent of the swirling region can be as large as $60 \xi$. 
This result suggests that a vortex-core region where the pressure can be lower than the surrounding region can be much larger than $\xi$.
We have also computed the curvature of vortex axes.
The PDF of the curvature of the vortex axes shows the power laws $\kappa^{1}$ in the small curvature region and $\kappa^{-5/2}$ in the large curvature region.
These vortex statistics, with a reference to the coherent length scale $\xi$, cannot be studied by the Biot-Savar model, and advantage of a GP simulation.
Among the statistical analysis, the PDF of vortex radius is reported for the first time,  

As above, our method is useful to elucidate various aspects of quantum vortices.
We are going to apply this method to GP simulations of developed quantum turbulence with a larger number of grid points.
We expect that an application of a vortex identification method, established in studies of classic fluid turbulence, to studies of GP turbulence can enable us to study quantum turbulence from various aspects, and highlight important physics which appears both in quantum and classic fluid turbulence.
Our next work will proceed including this point of view.

\begin{acknowledgments}
The author N. S. would like to take this opportunity to thank the ``Interdisciplinary Frontier Next-Generation Researcher Program of the Tokai Higher Education and Research System''
This research was partially supported by JSPS KAKENHI Grant Number 20H00225, Japan. 
The numerical simulations were performed on the Plasma Simulator of National Institute for Fusion Science (NEC SX-Aurora TSUBASA A412-8) with the support and under the auspices of the NIFS Collaboration Research program (NIFS20KNTS063), as well as on Wistria/BDEC-01 Odyssey (Fujitsu FX1000) supercomputer, being partially supported by the University of Tokyo through the HPCI System Research Project (Project ID: hp220036). 
\end{acknowledgments}


\begin{thebibliography}{67}%
\makeatletter
\providecommand \@ifxundefined [1]{%
 \@ifx{#1\undefined}
}%
\providecommand \@ifnum [1]{%
 \ifnum #1\expandafter \@firstoftwo
 \else \expandafter \@secondoftwo
 \fi
}%
\providecommand \@ifx [1]{%
 \ifx #1\expandafter \@firstoftwo
 \else \expandafter \@secondoftwo
 \fi
}%
\providecommand \natexlab [1]{#1}%
\providecommand \enquote  [1]{``#1''}%
\providecommand \bibnamefont  [1]{#1}%
\providecommand \bibfnamefont [1]{#1}%
\providecommand \citenamefont [1]{#1}%
\providecommand \href@noop [0]{\@secondoftwo}%
\providecommand \href [0]{\begingroup \@sanitize@url \@href}%
\providecommand \@href[1]{\@@startlink{#1}\@@href}%
\providecommand \@@href[1]{\endgroup#1\@@endlink}%
\providecommand \@sanitize@url [0]{\catcode `\\12\catcode `\$12\catcode
  `\&12\catcode `\#12\catcode `\^12\catcode `\_12\catcode `\%12\relax}%
\providecommand \@@startlink[1]{}%
\providecommand \@@endlink[0]{}%
\providecommand \url  [0]{\begingroup\@sanitize@url \@url }%
\providecommand \@url [1]{\endgroup\@href {#1}{\urlprefix }}%
\providecommand \urlprefix  [0]{URL }%
\providecommand \Eprint [0]{\href }%
\providecommand \doibase [0]{https://doi.org/}%
\providecommand \selectlanguage [0]{\@gobble}%
\providecommand \bibinfo  [0]{\@secondoftwo}%
\providecommand \bibfield  [0]{\@secondoftwo}%
\providecommand \translation [1]{[#1]}%
\providecommand \BibitemOpen [0]{}%
\providecommand \bibitemStop [0]{}%
\providecommand \bibitemNoStop [0]{.\EOS\space}%
\providecommand \EOS [0]{\spacefactor3000\relax}%
\providecommand \BibitemShut  [1]{\csname bibitem#1\endcsname}%
\let\auto@bib@innerbib\@empty
\bibitem [{\citenamefont {Vinen}\ and\ \citenamefont
  {Niemela}(2002)}]{Vinen2002}%
  \BibitemOpen
  \bibfield  {author} {\bibinfo {author} {\bibfnamefont {W.~F.}\ \bibnamefont
  {Vinen}}\ and\ \bibinfo {author} {\bibfnamefont {J.~J.}\ \bibnamefont
  {Niemela}},\ }\bibfield  {title} {\bibinfo {title} {Quantum turbulence},\
  }\href@noop {} {\bibfield  {journal} {\bibinfo  {journal} {J. Low Temp.
  Phys.}\ }\textbf {\bibinfo {volume} {128}},\ \bibinfo {pages} {167} (\bibinfo
  {year} {2002})}\BibitemShut {NoStop}%
\bibitem [{\citenamefont {Tsubota}(2008)}]{Tsubota2008}%
  \BibitemOpen
  \bibfield  {author} {\bibinfo {author} {\bibfnamefont {M.}~\bibnamefont
  {Tsubota}},\ }\bibfield  {title} {\bibinfo {title} {Quantum turbulence},\
  }\href {https://doi.org/10.1143/jpsj.77.111006} {\bibfield  {journal}
  {\bibinfo  {journal} {J. Phys. Soc. Jpn.}\ }\textbf {\bibinfo {volume}
  {77}},\ \bibinfo {pages} {111006} (\bibinfo {year} {2008})}\BibitemShut
  {NoStop}%
\bibitem [{\citenamefont {Paoletti}\ and\ \citenamefont
  {Lathrop}(2011)}]{Paoletti2011}%
  \BibitemOpen
  \bibfield  {author} {\bibinfo {author} {\bibfnamefont {M.~S.}\ \bibnamefont
  {Paoletti}}\ and\ \bibinfo {author} {\bibfnamefont {D.~P.}\ \bibnamefont
  {Lathrop}},\ }\bibfield  {title} {\bibinfo {title} {Quantum turbulence},\
  }\href {https://doi.org/10.1146/annurev-conmatphys-062910-140533} {\bibfield
  {journal} {\bibinfo  {journal} {Annu. Rev. Condens. Matter Phys.}\ }\textbf
  {\bibinfo {volume} {2}},\ \bibinfo {pages} {213} (\bibinfo {year}
  {2011})}\BibitemShut {NoStop}%
\bibitem [{\citenamefont {Saffman}(1992)}]{Saffman1992}%
  \BibitemOpen
  \bibfield  {author} {\bibinfo {author} {\bibfnamefont {P.}~\bibnamefont
  {Saffman}},\ }\href@noop {} {\emph {\bibinfo {title} {Vortex Dyanmics}}}\
  (\bibinfo  {publisher} {Cambridge University Press},\ \bibinfo {year}
  {1992})\BibitemShut {NoStop}%
\bibitem [{\citenamefont {Bao}\ and\ \citenamefont {Guo}(2019)}]{Bao2019}%
  \BibitemOpen
  \bibfield  {author} {\bibinfo {author} {\bibfnamefont {S.}~\bibnamefont
  {Bao}}\ and\ \bibinfo {author} {\bibfnamefont {W.}~\bibnamefont {Guo}},\
  }\bibfield  {title} {\bibinfo {title} {Quench-spot detection for
  superconducting accelerator cavities via flow visualization in superfluid
  helium-4},\ }\href@noop {} {\bibfield  {journal} {\bibinfo  {journal} {Phys.
  Rev. Appl.}\ }\textbf {\bibinfo {volume} {11}},\ \bibinfo {pages} {044003}
  (\bibinfo {year} {2019})}\BibitemShut {NoStop}%
\bibitem [{\citenamefont {Tatsumoto}\ \emph {et~al.}(2002)\citenamefont
  {Tatsumoto}, \citenamefont {Fukuda},\ and\ \citenamefont
  {Shiotsu}}]{Tatsumoto2002}%
  \BibitemOpen
  \bibfield  {author} {\bibinfo {author} {\bibfnamefont {H.}~\bibnamefont
  {Tatsumoto}}, \bibinfo {author} {\bibfnamefont {K.}~\bibnamefont {Fukuda}},\
  and\ \bibinfo {author} {\bibfnamefont {M.}~\bibnamefont {Shiotsu}},\
  }\bibfield  {title} {\bibinfo {title} {Numerical analysis for steady-state
  two-dimensional heat transfer from a flat plate at one side of a duct
  containing pressurized he ii},\ }\href
  {https://doi.org/https://doi.org/10.1016/S0011-2275(01)00155-2} {\bibfield
  {journal} {\bibinfo  {journal} {Cryogenics}\ }\textbf {\bibinfo {volume}
  {42}},\ \bibinfo {pages} {9} (\bibinfo {year} {2002})}\BibitemShut {NoStop}%
\bibitem [{\citenamefont {Bewley}\ \emph {et~al.}(2006)\citenamefont {Bewley},
  \citenamefont {Lathrop},\ and\ \citenamefont {Sreenivasan}}]{Bewley2006}%
  \BibitemOpen
  \bibfield  {author} {\bibinfo {author} {\bibfnamefont {G.~P.}\ \bibnamefont
  {Bewley}}, \bibinfo {author} {\bibfnamefont {D.~P.}\ \bibnamefont
  {Lathrop}},\ and\ \bibinfo {author} {\bibfnamefont {K.~R.}\ \bibnamefont
  {Sreenivasan}},\ }\bibfield  {title} {\bibinfo {title} {Visualization of
  quantized vortices},\ }\href@noop {} {\bibfield  {journal} {\bibinfo
  {journal} {Nature}\ }\textbf {\bibinfo {volume} {441}},\ \bibinfo {pages}
  {588} (\bibinfo {year} {2006})}\BibitemShut {NoStop}%
\bibitem [{\citenamefont {Paoletti}\ \emph {et~al.}(2008)\citenamefont
  {Paoletti}, \citenamefont {Fiorito}, \citenamefont {Sreenivasan},\ and\
  \citenamefont {Lathrop}}]{Paoletti2008}%
  \BibitemOpen
  \bibfield  {author} {\bibinfo {author} {\bibfnamefont {M.~S.}\ \bibnamefont
  {Paoletti}}, \bibinfo {author} {\bibfnamefont {R.~B.}\ \bibnamefont
  {Fiorito}}, \bibinfo {author} {\bibfnamefont {K.~R.}\ \bibnamefont
  {Sreenivasan}},\ and\ \bibinfo {author} {\bibfnamefont {D.~P.}\ \bibnamefont
  {Lathrop}},\ }\bibfield  {title} {\bibinfo {title} {Visualization of
  {S}uperfluid {H}elium {F}low},\ }\href
  {https://doi.org/10.1143/jpsj.77.111007} {\bibfield  {journal} {\bibinfo
  {journal} {J. Phys. Soc. Jpn.}\ }\textbf {\bibinfo {volume} {77}},\ \bibinfo
  {pages} {111007} (\bibinfo {year} {2008})}\BibitemShut {NoStop}%
\bibitem [{\citenamefont {Paoletti}\ \emph {et~al.}(2010)\citenamefont
  {Paoletti}, \citenamefont {Fisher},\ and\ \citenamefont
  {Lathrop}}]{Paoletti2010}%
  \BibitemOpen
  \bibfield  {author} {\bibinfo {author} {\bibfnamefont {M.}~\bibnamefont
  {Paoletti}}, \bibinfo {author} {\bibfnamefont {M.~E.}\ \bibnamefont
  {Fisher}},\ and\ \bibinfo {author} {\bibfnamefont {D.}~\bibnamefont
  {Lathrop}},\ }\bibfield  {title} {\bibinfo {title} {Reconnection dynamics for
  quantized vortices},\ }\href@noop {} {\bibfield  {journal} {\bibinfo
  {journal} {Physica D}\ }\textbf {\bibinfo {volume} {239}},\ \bibinfo {pages}
  {1367} (\bibinfo {year} {2010})}\BibitemShut {NoStop}%
\bibitem [{\citenamefont {Chagovets}\ and\ \citenamefont
  {Van~Sciver}(2011)}]{Chagovets2011}%
  \BibitemOpen
  \bibfield  {author} {\bibinfo {author} {\bibfnamefont {T.~V.}\ \bibnamefont
  {Chagovets}}\ and\ \bibinfo {author} {\bibfnamefont {S.~W.}\ \bibnamefont
  {Van~Sciver}},\ }\bibfield  {title} {\bibinfo {title} {A study of thermal
  counterflow using particle tracking velocimetry},\ }\href@noop {} {\bibfield
  {journal} {\bibinfo  {journal} {Phys. Fluids}\ }\textbf {\bibinfo {volume}
  {23}},\ \bibinfo {pages} {107102} (\bibinfo {year} {2011})}\BibitemShut
  {NoStop}%
\bibitem [{\citenamefont {La~Mantia}\ \emph {et~al.}(2013)\citenamefont
  {La~Mantia}, \citenamefont {Duda}, \citenamefont {Rotter},\ and\
  \citenamefont {Skrbek}}]{Mantia2013}%
  \BibitemOpen
  \bibfield  {author} {\bibinfo {author} {\bibfnamefont {M.}~\bibnamefont
  {La~Mantia}}, \bibinfo {author} {\bibfnamefont {D.}~\bibnamefont {Duda}},
  \bibinfo {author} {\bibfnamefont {M.}~\bibnamefont {Rotter}},\ and\ \bibinfo
  {author} {\bibfnamefont {L.}~\bibnamefont {Skrbek}},\ }\bibfield  {title}
  {\bibinfo {title} {Lagrangian accelerations of particles in superfluid
  turbulence},\ }\href {https://doi.org/10.1017/jfm.2013.31} {\bibfield
  {journal} {\bibinfo  {journal} {J. Fluid Mech.}\ }\textbf {\bibinfo {volume}
  {717}},\ \bibinfo {pages} {R9} (\bibinfo {year} {2013})}\BibitemShut
  {NoStop}%
\bibitem [{\citenamefont {Mantia}\ and\ \citenamefont
  {Skrbek}(2014)}]{Mantia2014a}%
  \BibitemOpen
  \bibfield  {author} {\bibinfo {author} {\bibfnamefont {M.~L.}\ \bibnamefont
  {Mantia}}\ and\ \bibinfo {author} {\bibfnamefont {L.}~\bibnamefont
  {Skrbek}},\ }\bibfield  {title} {\bibinfo {title} {Quantum, or classical
  turbulence?},\ }\href@noop {} {\bibfield  {journal} {\bibinfo  {journal}
  {EPL}\ }\textbf {\bibinfo {volume} {105}},\ \bibinfo {pages} {46002}
  (\bibinfo {year} {2014})}\BibitemShut {NoStop}%
\bibitem [{\citenamefont {La~Mantia}\ and\ \citenamefont
  {Skrbek}(2014)}]{Mantia2014b}%
  \BibitemOpen
  \bibfield  {author} {\bibinfo {author} {\bibfnamefont {M.}~\bibnamefont
  {La~Mantia}}\ and\ \bibinfo {author} {\bibfnamefont {L.}~\bibnamefont
  {Skrbek}},\ }\bibfield  {title} {\bibinfo {title} {Quantum turbulence
  visualized by particle dynamics},\ }\href@noop {} {\bibfield  {journal}
  {\bibinfo  {journal} {Phys. Rev. B}\ }\textbf {\bibinfo {volume} {90}},\
  \bibinfo {pages} {014519} (\bibinfo {year} {2014})}\BibitemShut {NoStop}%
\bibitem [{\citenamefont {La~Mantia}(2016)}]{Mantia2016}%
  \BibitemOpen
  \bibfield  {author} {\bibinfo {author} {\bibfnamefont {M.}~\bibnamefont
  {La~Mantia}},\ }\bibfield  {title} {\bibinfo {title} {Particle trajectories
  in thermal counterflow of superfluid helium in a wide channel of square cross
  section},\ }\href@noop {} {\bibfield  {journal} {\bibinfo  {journal} {Phys.
  Fluids}\ }\textbf {\bibinfo {volume} {28}},\ \bibinfo {pages} {024102}
  (\bibinfo {year} {2016})}\BibitemShut {NoStop}%
\bibitem [{\citenamefont {Kubo}\ and\ \citenamefont {Tsuji}(2017)}]{Kubo2017}%
  \BibitemOpen
  \bibfield  {author} {\bibinfo {author} {\bibfnamefont {W.}~\bibnamefont
  {Kubo}}\ and\ \bibinfo {author} {\bibfnamefont {Y.}~\bibnamefont {Tsuji}},\
  }\bibfield  {title} {\bibinfo {title} {Lagrangian {T}rajectory of {S}mall
  {P}articles in {S}uperfluid {H}e {II}},\ }\href@noop {} {\bibfield  {journal}
  {\bibinfo  {journal} {J. Low Temp. Phys.}\ }\textbf {\bibinfo {volume}
  {187}},\ \bibinfo {pages} {611} (\bibinfo {year} {2017})}\BibitemShut
  {NoStop}%
\bibitem [{\citenamefont {Mastracci}\ and\ \citenamefont
  {Guo}(2018)}]{Mastracci2018}%
  \BibitemOpen
  \bibfield  {author} {\bibinfo {author} {\bibfnamefont {B.}~\bibnamefont
  {Mastracci}}\ and\ \bibinfo {author} {\bibfnamefont {W.}~\bibnamefont
  {Guo}},\ }\bibfield  {title} {\bibinfo {title} {Exploration of thermal
  counterflow in {H}e {II} using particle tracking velocimetry},\ }\href@noop
  {} {\bibfield  {journal} {\bibinfo  {journal} {Phys. Fluids}\ }\textbf
  {\bibinfo {volume} {3}},\ \bibinfo {pages} {063304} (\bibinfo {year}
  {2018})}\BibitemShut {NoStop}%
\bibitem [{\citenamefont {Kubo}\ and\ \citenamefont {Tsuji}(2019)}]{Kubo2019}%
  \BibitemOpen
  \bibfield  {author} {\bibinfo {author} {\bibfnamefont {W.}~\bibnamefont
  {Kubo}}\ and\ \bibinfo {author} {\bibfnamefont {Y.}~\bibnamefont {Tsuji}},\
  }\bibfield  {title} {\bibinfo {title} {{S}tatistical {P}roperties of {S}mall
  {P}article {T}rajectories in a {F}ully {D}eveloped {T}urbulent {S}tate in
  {H}e-{II}},\ }\href@noop {} {\bibfield  {journal} {\bibinfo  {journal} {J.
  Low Temp. Phys.}\ }\textbf {\bibinfo {volume} {196}},\ \bibinfo {pages} {170}
  (\bibinfo {year} {2019})}\BibitemShut {NoStop}%
\bibitem [{\citenamefont {Tang}\ \emph {et~al.}(2021)\citenamefont {Tang},
  \citenamefont {Bao},\ and\ \citenamefont {Guo}}]{Tang2021}%
  \BibitemOpen
  \bibfield  {author} {\bibinfo {author} {\bibfnamefont {Y.}~\bibnamefont
  {Tang}}, \bibinfo {author} {\bibfnamefont {S.}~\bibnamefont {Bao}},\ and\
  \bibinfo {author} {\bibfnamefont {W.}~\bibnamefont {Guo}},\ }\bibfield
  {title} {\bibinfo {title} {Superdiffusion of quantized vortices uncovering
  scaling laws in quantum turbulence},\ }\href@noop {} {\bibfield  {journal}
  {\bibinfo  {journal} {Proc. Natl. Acad. Sci. U.S.A.}\ }\textbf {\bibinfo
  {volume} {118}},\ \bibinfo {pages} {e2021957118} (\bibinfo {year}
  {2021})}\BibitemShut {NoStop}%
\bibitem [{\citenamefont {Svan\v{c}ara}\ \emph {et~al.}(2021)\citenamefont
  {Svan\v{c}ara}, \citenamefont {Duda}, \citenamefont {Hrubcov\'{a}},
  \citenamefont {Rotter}, \citenamefont {Skrbek}, \citenamefont {La~Mantia},
  \citenamefont {Durozoy}, \citenamefont {Diribarne}, \citenamefont {Rousset},
  \citenamefont {Bourgoin},\ and\ \citenamefont {et~al.}}]{Svancara2021}%
  \BibitemOpen
  \bibfield  {author} {\bibinfo {author} {\bibfnamefont {P.}~\bibnamefont
  {Svan\v{c}ara}}, \bibinfo {author} {\bibfnamefont {D.}~\bibnamefont {Duda}},
  \bibinfo {author} {\bibfnamefont {P.}~\bibnamefont {Hrubcov\'{a}}}, \bibinfo
  {author} {\bibfnamefont {M.}~\bibnamefont {Rotter}}, \bibinfo {author}
  {\bibfnamefont {L.}~\bibnamefont {Skrbek}}, \bibinfo {author} {\bibfnamefont
  {M.}~\bibnamefont {La~Mantia}}, \bibinfo {author} {\bibfnamefont
  {E.}~\bibnamefont {Durozoy}}, \bibinfo {author} {\bibfnamefont
  {P.}~\bibnamefont {Diribarne}}, \bibinfo {author} {\bibfnamefont
  {B.}~\bibnamefont {Rousset}}, \bibinfo {author} {\bibfnamefont
  {M.}~\bibnamefont {Bourgoin}},\ and\ \bibinfo {author} {\bibnamefont
  {et~al.}},\ }\bibfield  {title} {\bibinfo {title} {Ubiquity of
  particle$-$vortex interactions in turbulent counterflow of superfluid
  helium},\ }\href {https://doi.org/10.1017/jfm.2020.1017} {\bibfield
  {journal} {\bibinfo  {journal} {J. Fluid Mech.}\ }\textbf {\bibinfo {volume}
  {911}},\ \bibinfo {pages} {A8} (\bibinfo {year} {2021})}\BibitemShut
  {NoStop}%
\bibitem [{\citenamefont {Chen}\ \emph {et~al.}(2022)\citenamefont {Chen},
  \citenamefont {Maruyama},\ and\ \citenamefont {Tsuji}}]{Chen2022a}%
  \BibitemOpen
  \bibfield  {author} {\bibinfo {author} {\bibfnamefont {L.}~\bibnamefont
  {Chen}}, \bibinfo {author} {\bibfnamefont {T.}~\bibnamefont {Maruyama}},\
  and\ \bibinfo {author} {\bibfnamefont {Y.}~\bibnamefont {Tsuji}},\ }\bibfield
   {title} {\bibinfo {title} {Statistical properties of lagrangian trajectories
  of small particles in superfluid $^4$he},\ }\href@noop {} {\bibfield
  {journal} {\bibinfo  {journal} {J. Low Temp. Phys.}\ }\textbf {\bibinfo
  {volume} {208}},\ \bibinfo {pages} {402} (\bibinfo {year}
  {2022})}\BibitemShut {NoStop}%
\bibitem [{\citenamefont {Chen}\ and\ \citenamefont {Tsuji}(2022)}]{Chen2022b}%
  \BibitemOpen
  \bibfield  {author} {\bibinfo {author} {\bibfnamefont {L.}~\bibnamefont
  {Chen}}\ and\ \bibinfo {author} {\bibfnamefont {Y.}~\bibnamefont {Tsuji}},\
  }\bibfield  {title} {\bibinfo {title} {Trajectory analysis of particle
  motions in superfluid helium-4 using ptv method},\ }\href@noop {} {\bibfield
  {journal} {\bibinfo  {journal} {J. Flow Control. Meas. Vis.}\ }\textbf
  {\bibinfo {volume} {10}},\ \bibinfo {pages} {76} (\bibinfo {year}
  {2022})}\BibitemShut {NoStop}%
\bibitem [{\citenamefont {Sakaki}\ \emph
  {et~al.}(2022{\natexlab{a}})\citenamefont {Sakaki}, \citenamefont
  {Maruyama},\ and\ \citenamefont {Tsuji}}]{Sakaki2022a}%
  \BibitemOpen
  \bibfield  {author} {\bibinfo {author} {\bibfnamefont {N.}~\bibnamefont
  {Sakaki}}, \bibinfo {author} {\bibfnamefont {T.}~\bibnamefont {Maruyama}},\
  and\ \bibinfo {author} {\bibfnamefont {Y.}~\bibnamefont {Tsuji}},\ }\bibfield
   {title} {\bibinfo {title} {Statistics of the {L}agrangian {T}rajectories'
  {C}urvature in {T}hermal {C}ounterflow},\ }\href@noop {} {\bibfield
  {journal} {\bibinfo  {journal} {J. Low Temp. Phys.}\ }\textbf {\bibinfo
  {volume} {208}},\ \bibinfo {pages} {418} (\bibinfo {year}
  {2022}{\natexlab{a}})}\BibitemShut {NoStop}%
\bibitem [{\citenamefont {Sakaki}\ \emph
  {et~al.}(2022{\natexlab{b}})\citenamefont {Sakaki}, \citenamefont
  {Maruyama},\ and\ \citenamefont {Tsuji}}]{Sakaki2022b}%
  \BibitemOpen
  \bibfield  {author} {\bibinfo {author} {\bibfnamefont {N.}~\bibnamefont
  {Sakaki}}, \bibinfo {author} {\bibfnamefont {T.}~\bibnamefont {Maruyama}},\
  and\ \bibinfo {author} {\bibfnamefont {Y.}~\bibnamefont {Tsuji}},\ }\bibfield
   {title} {\bibinfo {title} {Study on the curvature of lagrangian trajectories
  in thermal counterflow},\ }\href@noop {} {\bibfield  {journal} {\bibinfo
  {journal} {J. Low Temp. Phys.}\ }\textbf {\bibinfo {volume} {208}},\ \bibinfo
  {pages} {223} (\bibinfo {year} {2022}{\natexlab{b}})}\BibitemShut {NoStop}%
\bibitem [{\citenamefont {Schwarz}(1985)}]{Schwarz1985}%
  \BibitemOpen
  \bibfield  {author} {\bibinfo {author} {\bibfnamefont {K.~W.}\ \bibnamefont
  {Schwarz}},\ }\bibfield  {title} {\bibinfo {title} {Three-dimensional vortex
  dynamics in superfluid $^{4}\mathrm{He}$: Line-line and line-boundary
  interactions},\ }\href@noop {} {\bibfield  {journal} {\bibinfo  {journal}
  {Phys. Rev. B}\ }\textbf {\bibinfo {volume} {31}},\ \bibinfo {pages} {5782}
  (\bibinfo {year} {1985})}\BibitemShut {NoStop}%
\bibitem [{\citenamefont {Adachi}\ \emph {et~al.}(2010)\citenamefont {Adachi},
  \citenamefont {Fujiyama},\ and\ \citenamefont {Tsubota}}]{Adachi2010}%
  \BibitemOpen
  \bibfield  {author} {\bibinfo {author} {\bibfnamefont {H.}~\bibnamefont
  {Adachi}}, \bibinfo {author} {\bibfnamefont {S.}~\bibnamefont {Fujiyama}},\
  and\ \bibinfo {author} {\bibfnamefont {M.}~\bibnamefont {Tsubota}},\
  }\bibfield  {title} {\bibinfo {title} {Steady-state counterflow quantum
  turbulence: Simulation of vortex filaments using the full biot-savart law},\
  }\href@noop {} {\bibfield  {journal} {\bibinfo  {journal} {Phys. Rev. B}\
  }\textbf {\bibinfo {volume} {81}},\ \bibinfo {pages} {104511} (\bibinfo
  {year} {2010})}\BibitemShut {NoStop}%
\bibitem [{\citenamefont {Bou\'e}\ \emph {et~al.}(2013)\citenamefont {Bou\'e},
  \citenamefont {Khomenko}, \citenamefont {L'vov},\ and\ \citenamefont
  {Procaccia}}]{Boue2013}%
  \BibitemOpen
  \bibfield  {author} {\bibinfo {author} {\bibfnamefont {L.}~\bibnamefont
  {Bou\'e}}, \bibinfo {author} {\bibfnamefont {D.}~\bibnamefont {Khomenko}},
  \bibinfo {author} {\bibfnamefont {V.~S.}\ \bibnamefont {L'vov}},\ and\
  \bibinfo {author} {\bibfnamefont {I.}~\bibnamefont {Procaccia}},\ }\bibfield
  {title} {\bibinfo {title} {Analytic solution of the approach of quantum
  vortices towards reconnection},\ }\href@noop {} {\bibfield  {journal}
  {\bibinfo  {journal} {Phys. Rev. Lett.}\ }\textbf {\bibinfo {volume} {111}},\
  \bibinfo {pages} {145302} (\bibinfo {year} {2013})}\BibitemShut {NoStop}%
\bibitem [{\citenamefont {Mineda}\ \emph {et~al.}(2013)\citenamefont {Mineda},
  \citenamefont {Tsubota}, \citenamefont {Sergeev}, \citenamefont {Barenghi},\
  and\ \citenamefont {Vinen}}]{Mineda2013}%
  \BibitemOpen
  \bibfield  {author} {\bibinfo {author} {\bibfnamefont {Y.}~\bibnamefont
  {Mineda}}, \bibinfo {author} {\bibfnamefont {M.}~\bibnamefont {Tsubota}},
  \bibinfo {author} {\bibfnamefont {Y.~A.}\ \bibnamefont {Sergeev}}, \bibinfo
  {author} {\bibfnamefont {C.~F.}\ \bibnamefont {Barenghi}},\ and\ \bibinfo
  {author} {\bibfnamefont {W.~F.}\ \bibnamefont {Vinen}},\ }\bibfield  {title}
  {\bibinfo {title} {Velocity distributions of tracer particles in thermal
  counterflow in superfluid ${}^{4}$he},\ }\href@noop {} {\bibfield  {journal}
  {\bibinfo  {journal} {Phys. Rev. B}\ }\textbf {\bibinfo {volume} {87}},\
  \bibinfo {pages} {174508} (\bibinfo {year} {2013})}\BibitemShut {NoStop}%
\bibitem [{\citenamefont {Yui}\ and\ \citenamefont {Tsubota}(2015)}]{Yui2015}%
  \BibitemOpen
  \bibfield  {author} {\bibinfo {author} {\bibfnamefont {S.}~\bibnamefont
  {Yui}}\ and\ \bibinfo {author} {\bibfnamefont {M.}~\bibnamefont {Tsubota}},\
  }\bibfield  {title} {\bibinfo {title} {Counterflow quantum turbulence of
  he-ii in a square channel: Numerical analysis with nonuniform flows of the
  normal fluid},\ }\href@noop {} {\bibfield  {journal} {\bibinfo  {journal}
  {Phys. Rev. B}\ }\textbf {\bibinfo {volume} {91}},\ \bibinfo {pages} {184504}
  (\bibinfo {year} {2015})}\BibitemShut {NoStop}%
\bibitem [{\citenamefont {Yui}\ \emph {et~al.}(2020)\citenamefont {Yui},
  \citenamefont {Kobayashi}, \citenamefont {Tsubota},\ and\ \citenamefont
  {Guo}}]{Yui2020}%
  \BibitemOpen
  \bibfield  {author} {\bibinfo {author} {\bibfnamefont {S.}~\bibnamefont
  {Yui}}, \bibinfo {author} {\bibfnamefont {H.}~\bibnamefont {Kobayashi}},
  \bibinfo {author} {\bibfnamefont {M.}~\bibnamefont {Tsubota}},\ and\ \bibinfo
  {author} {\bibfnamefont {W.}~\bibnamefont {Guo}},\ }\bibfield  {title}
  {\bibinfo {title} {Fully coupled two-fluid dynamics in superfluid
  $^{4}\mathrm{He}$: Anomalous anisotropic velocity fluctuations in
  counterflow},\ }\href@noop {} {\bibfield  {journal} {\bibinfo  {journal}
  {Phys. Rev. Lett.}\ }\textbf {\bibinfo {volume} {124}},\ \bibinfo {pages}
  {155301} (\bibinfo {year} {2020})}\BibitemShut {NoStop}%
\bibitem [{\citenamefont {Nore}\ \emph {et~al.}(1997)\citenamefont {Nore},
  \citenamefont {Abid},\ and\ \citenamefont {Brachet}}]{Nore1997}%
  \BibitemOpen
  \bibfield  {author} {\bibinfo {author} {\bibfnamefont {C.}~\bibnamefont
  {Nore}}, \bibinfo {author} {\bibfnamefont {M.}~\bibnamefont {Abid}},\ and\
  \bibinfo {author} {\bibfnamefont {M.~E.}\ \bibnamefont {Brachet}},\
  }\bibfield  {title} {\bibinfo {title} {Kolmogorov turbulence in
  low-temperature superflows},\ }\href@noop {} {\bibfield  {journal} {\bibinfo
  {journal} {Phys. Rev. Lett.}\ }\textbf {\bibinfo {volume} {78}},\ \bibinfo
  {pages} {3896} (\bibinfo {year} {1997})}\BibitemShut {NoStop}%
\bibitem [{\citenamefont {Kobayashi}\ and\ \citenamefont
  {Tsubota}(2005)}]{Kobayashi2005}%
  \BibitemOpen
  \bibfield  {author} {\bibinfo {author} {\bibfnamefont {M.}~\bibnamefont
  {Kobayashi}}\ and\ \bibinfo {author} {\bibfnamefont {M.}~\bibnamefont
  {Tsubota}},\ }\bibfield  {title} {\bibinfo {title} {Kolmogorov spectrum of
  superfluid turbulence: Numerical analysis of the gross-pitaevskii equation
  with a small-scale dissipation},\ }\href@noop {} {\bibfield  {journal}
  {\bibinfo  {journal} {Phys. Rev. Lett.}\ }\textbf {\bibinfo {volume} {94}},\
  \bibinfo {pages} {065302} (\bibinfo {year} {2005})}\BibitemShut {NoStop}%
\bibitem [{\citenamefont {Kobayashi}\ and\ \citenamefont
  {Tsubota}(2006)}]{Kobayashi2006}%
  \BibitemOpen
  \bibfield  {author} {\bibinfo {author} {\bibfnamefont {M.}~\bibnamefont
  {Kobayashi}}\ and\ \bibinfo {author} {\bibfnamefont {M.}~\bibnamefont
  {Tsubota}},\ }\bibfield  {title} {\bibinfo {title} {Decay of quantized
  vortices in quantum turbulence},\ }\href@noop {} {\bibfield  {journal}
  {\bibinfo  {journal} {J. Low Temp. Jpn.}\ }\textbf {\bibinfo {volume}
  {145}},\ \bibinfo {pages} {209} (\bibinfo {year} {2006})}\BibitemShut
  {NoStop}%
\bibitem [{\citenamefont {Zuccher}\ \emph {et~al.}(2012)\citenamefont
  {Zuccher}, \citenamefont {Caliari}, \citenamefont {Baggaley},\ and\
  \citenamefont {Barenghi}}]{Zuccher2012}%
  \BibitemOpen
  \bibfield  {author} {\bibinfo {author} {\bibfnamefont {S.}~\bibnamefont
  {Zuccher}}, \bibinfo {author} {\bibfnamefont {M.}~\bibnamefont {Caliari}},
  \bibinfo {author} {\bibfnamefont {A.~W.}\ \bibnamefont {Baggaley}},\ and\
  \bibinfo {author} {\bibfnamefont {C.~F.}\ \bibnamefont {Barenghi}},\
  }\bibfield  {title} {\bibinfo {title} {Quantum vortex reconnections},\
  }\href@noop {} {\bibfield  {journal} {\bibinfo  {journal} {Phys. Fluids}\
  }\textbf {\bibinfo {volume} {24}},\ \bibinfo {pages} {125108} (\bibinfo
  {year} {2012})}\BibitemShut {NoStop}%
\bibitem [{\citenamefont {Villois}\ \emph
  {et~al.}(2016{\natexlab{a}})\citenamefont {Villois}, \citenamefont
  {Proment},\ and\ \citenamefont {Krstulovic}}]{Villois2016b}%
  \BibitemOpen
  \bibfield  {author} {\bibinfo {author} {\bibfnamefont {A.}~\bibnamefont
  {Villois}}, \bibinfo {author} {\bibfnamefont {D.}~\bibnamefont {Proment}},\
  and\ \bibinfo {author} {\bibfnamefont {G.}~\bibnamefont {Krstulovic}},\
  }\bibfield  {title} {\bibinfo {title} {Evolution of a superfluid vortex
  filament tangle driven by the gross-pitaevskii equation},\ }\href@noop {}
  {\bibfield  {journal} {\bibinfo  {journal} {Phys. Rev. E}\ }\textbf {\bibinfo
  {volume} {93}},\ \bibinfo {pages} {061103} (\bibinfo {year}
  {2016}{\natexlab{a}})}\BibitemShut {NoStop}%
\bibitem [{\citenamefont {Stagg}\ \emph {et~al.}(2017)\citenamefont {Stagg},
  \citenamefont {Parker},\ and\ \citenamefont {Barenghi}}]{Stagg2017}%
  \BibitemOpen
  \bibfield  {author} {\bibinfo {author} {\bibfnamefont {G.~W.}\ \bibnamefont
  {Stagg}}, \bibinfo {author} {\bibfnamefont {N.~G.}\ \bibnamefont {Parker}},\
  and\ \bibinfo {author} {\bibfnamefont {C.~F.}\ \bibnamefont {Barenghi}},\
  }\bibfield  {title} {\bibinfo {title} {Superfluid boundary layer},\
  }\href@noop {} {\bibfield  {journal} {\bibinfo  {journal} {Phys. Rev. Lett.}\
  }\textbf {\bibinfo {volume} {118}},\ \bibinfo {pages} {135301} (\bibinfo
  {year} {2017})}\BibitemShut {NoStop}%
\bibitem [{\citenamefont {Yoshida}\ \emph {et~al.}(2019)\citenamefont
  {Yoshida}, \citenamefont {Miura},\ and\ \citenamefont {Tsuji}}]{Yoshida2019}%
  \BibitemOpen
  \bibfield  {author} {\bibinfo {author} {\bibfnamefont {K.}~\bibnamefont
  {Yoshida}}, \bibinfo {author} {\bibfnamefont {H.}~\bibnamefont {Miura}},\
  and\ \bibinfo {author} {\bibfnamefont {Y.}~\bibnamefont {Tsuji}},\ }\bibfield
   {title} {\bibinfo {title} {Spectrum in the strong turbulence region of
  gross--pitaevskii turbulence},\ }\href@noop {} {\bibfield  {journal}
  {\bibinfo  {journal} {J. Low Temp. Phys.}\ }\textbf {\bibinfo {volume}
  {196}},\ \bibinfo {pages} {211} (\bibinfo {year} {2019})}\BibitemShut
  {NoStop}%
\bibitem [{\citenamefont {Berloff}\ and\ \citenamefont
  {Youd}(2007)}]{Berloff2007}%
  \BibitemOpen
  \bibfield  {author} {\bibinfo {author} {\bibfnamefont {N.~G.}\ \bibnamefont
  {Berloff}}\ and\ \bibinfo {author} {\bibfnamefont {A.~J.}\ \bibnamefont
  {Youd}},\ }\bibfield  {title} {\bibinfo {title} {Dissipative dynamics of
  superfluid vortices at nonzero temperatures},\ }\href
  {https://doi.org/10.1103/PhysRevLett.99.145301} {\bibfield  {journal}
  {\bibinfo  {journal} {Phys. Rev. Lett.}\ }\textbf {\bibinfo {volume} {99}},\
  \bibinfo {pages} {145301} (\bibinfo {year} {2007})}\BibitemShut {NoStop}%
\bibitem [{\citenamefont {White}\ \emph {et~al.}(2010)\citenamefont {White},
  \citenamefont {Barenghi}, \citenamefont {Proukakis}, \citenamefont {Youd},\
  and\ \citenamefont {Wacks}}]{White2010}%
  \BibitemOpen
  \bibfield  {author} {\bibinfo {author} {\bibfnamefont {A.~C.}\ \bibnamefont
  {White}}, \bibinfo {author} {\bibfnamefont {C.~F.}\ \bibnamefont {Barenghi}},
  \bibinfo {author} {\bibfnamefont {N.~P.}\ \bibnamefont {Proukakis}}, \bibinfo
  {author} {\bibfnamefont {A.~J.}\ \bibnamefont {Youd}},\ and\ \bibinfo
  {author} {\bibfnamefont {D.~H.}\ \bibnamefont {Wacks}},\ }\bibfield  {title}
  {\bibinfo {title} {Nonclassical velocity statistics in a turbulent atomic
  bose-einstein condensate},\ }\href
  {https://doi.org/10.1103/PhysRevLett.104.075301} {\bibfield  {journal}
  {\bibinfo  {journal} {Phys. Rev. Lett.}\ }\textbf {\bibinfo {volume} {104}},\
  \bibinfo {pages} {075301} (\bibinfo {year} {2010})}\BibitemShut {NoStop}%
\bibitem [{\citenamefont {Villois}\ \emph
  {et~al.}(2016{\natexlab{b}})\citenamefont {Villois}, \citenamefont
  {Krstulovic}, \citenamefont {Proment},\ and\ \citenamefont
  {Salman}}]{Villois2016}%
  \BibitemOpen
  \bibfield  {author} {\bibinfo {author} {\bibfnamefont {A.}~\bibnamefont
  {Villois}}, \bibinfo {author} {\bibfnamefont {G.}~\bibnamefont {Krstulovic}},
  \bibinfo {author} {\bibfnamefont {D.}~\bibnamefont {Proment}},\ and\ \bibinfo
  {author} {\bibfnamefont {H.}~\bibnamefont {Salman}},\ }\bibfield  {title}
  {\bibinfo {title} {A vortex filament tracking method for the
  gross{\textendash}pitaevskii model of a superfluid},\ }\href@noop {}
  {\bibfield  {journal} {\bibinfo  {journal} {J. Phys. A: Math. Theor.}\
  }\textbf {\bibinfo {volume} {49}},\ \bibinfo {pages} {415502} (\bibinfo
  {year} {2016}{\natexlab{b}})}\BibitemShut {NoStop}%
\bibitem [{\citenamefont {Yoshida}\ \emph {et~al.}(2023)\citenamefont
  {Yoshida}, \citenamefont {Miura},\ and\ \citenamefont {Tsuji}}]{Yoshida2022}%
  \BibitemOpen
  \bibfield  {author} {\bibinfo {author} {\bibfnamefont {K.}~\bibnamefont
  {Yoshida}}, \bibinfo {author} {\bibfnamefont {H.}~\bibnamefont {Miura}},\
  and\ \bibinfo {author} {\bibfnamefont {Y.}~\bibnamefont {Tsuji}},\ }\bibfield
   {title} {\bibinfo {title} {Energy transfer of the gross-pitaevskii
  turbulence in weak-wave-turbulence and strong-turbulence ranges},\
  }\href@noop {} {\bibfield  {journal} {\bibinfo  {journal} {J. Low Temp.
  Phys.}\ }\textbf {\bibinfo {volume} {210}},\ \bibinfo {pages} {103} (\bibinfo
  {year} {2023})}\BibitemShut {NoStop}%
\bibitem [{\citenamefont {Hussain}\ and\ \citenamefont
  {Duraisamy}(2011)}]{HussainDuraisamy2021}%
  \BibitemOpen
  \bibfield  {author} {\bibinfo {author} {\bibfnamefont {F.}~\bibnamefont
  {Hussain}}\ and\ \bibinfo {author} {\bibfnamefont {K.}~\bibnamefont
  {Duraisamy}},\ }\bibfield  {title} {\bibinfo {title} {Mechanics of viscous
  vortex reconnection},\ }\href {https://doi.org/10.1063/1.3532039} {\bibfield
  {journal} {\bibinfo  {journal} {Phys. Fluids}\ }\textbf {\bibinfo {volume}
  {23}},\ \bibinfo {pages} {021701} (\bibinfo {year} {2011})}\BibitemShut
  {NoStop}%
\bibitem [{\citenamefont {Taylor}\ and\ \citenamefont
  {Dennis}(2014)}]{Taylor2014}%
  \BibitemOpen
  \bibfield  {author} {\bibinfo {author} {\bibfnamefont {A.~J.}\ \bibnamefont
  {Taylor}}\ and\ \bibinfo {author} {\bibfnamefont {M.~R.}\ \bibnamefont
  {Dennis}},\ }\bibfield  {title} {\bibinfo {title} {Geometry and scaling of
  tangled vortex lines in three-dimensional random wave fields},\ }\href
  {https://doi.org/10.1088/1751-8113/47/46/465101} {\bibfield  {journal}
  {\bibinfo  {journal} {J. Phys. A: Math. Theor.}\ }\textbf {\bibinfo {volume}
  {47}},\ \bibinfo {pages} {465101} (\bibinfo {year} {2014})}\BibitemShut
  {NoStop}%
\bibitem [{\citenamefont {Proment}\ \emph {et~al.}(2013)\citenamefont
  {Proment}, \citenamefont {Barenghi},\ and\ \citenamefont
  {Onorato}}]{Proment2013}%
  \BibitemOpen
  \bibfield  {author} {\bibinfo {author} {\bibfnamefont {D.}~\bibnamefont
  {Proment}}, \bibinfo {author} {\bibfnamefont {C.~F.}\ \bibnamefont
  {Barenghi}},\ and\ \bibinfo {author} {\bibfnamefont {M.}~\bibnamefont
  {Onorato}},\ }\href {https://doi.org/10.48550/ARXIV.1308.0852} {\bibinfo
  {title} {10.48550/arxiv.1308.0852}} (\bibinfo {year} {2013})\BibitemShut
  {NoStop}%
\bibitem [{\citenamefont {Krstulovic}(2012)}]{Krstulovic2012}%
  \BibitemOpen
  \bibfield  {author} {\bibinfo {author} {\bibfnamefont {G.}~\bibnamefont
  {Krstulovic}},\ }\bibfield  {title} {\bibinfo {title} {Kelvin-wave cascade
  and dissipation in low-temperature superfluid vortices},\ }\href@noop {}
  {\bibfield  {journal} {\bibinfo  {journal} {Phys. Rev. E}\ }\textbf {\bibinfo
  {volume} {86}},\ \bibinfo {pages} {055301(R)} (\bibinfo {year}
  {2012})}\BibitemShut {NoStop}%
\bibitem [{\citenamefont {Miura}\ and\ \citenamefont {Kida}(1997)}]{Miura1997}%
  \BibitemOpen
  \bibfield  {author} {\bibinfo {author} {\bibfnamefont {H.}~\bibnamefont
  {Miura}}\ and\ \bibinfo {author} {\bibfnamefont {S.}~\bibnamefont {Kida}},\
  }\bibfield  {title} {\bibinfo {title} {Identification of tubular vortices in
  turbulence},\ }\href@noop {} {\bibfield  {journal} {\bibinfo  {journal} {J.
  Phys. Soc. Jpn.}\ }\textbf {\bibinfo {volume} {66}},\ \bibinfo {pages} {1331}
  (\bibinfo {year} {1997})}\BibitemShut {NoStop}%
\bibitem [{\citenamefont {Kida}\ and\ \citenamefont
  {Miura}(1998{\natexlab{a}})}]{Kida1998}%
  \BibitemOpen
  \bibfield  {author} {\bibinfo {author} {\bibfnamefont {S.}~\bibnamefont
  {Kida}}\ and\ \bibinfo {author} {\bibfnamefont {H.}~\bibnamefont {Miura}},\
  }\bibfield  {title} {\bibinfo {title} {Swirl condition in low-pressure
  vortices},\ }\href@noop {} {\bibfield  {journal} {\bibinfo  {journal} {J.
  Phys. Soc. Jpn.}\ }\textbf {\bibinfo {volume} {67}},\ \bibinfo {pages} {2166}
  (\bibinfo {year} {1998}{\natexlab{a}})}\BibitemShut {NoStop}%
\bibitem [{\citenamefont {Kida}\ and\ \citenamefont
  {Miura}(1998{\natexlab{b}})}]{Kida1998b}%
  \BibitemOpen
  \bibfield  {author} {\bibinfo {author} {\bibfnamefont {S.}~\bibnamefont
  {Kida}}\ and\ \bibinfo {author} {\bibfnamefont {H.}~\bibnamefont {Miura}},\
  }\bibfield  {title} {\bibinfo {title} {Identification and analysis of
  vortical structures},\ }\href@noop {} {\bibfield  {journal} {\bibinfo
  {journal} {Euro. J. Mech. B/Fluids}\ }\textbf {\bibinfo {volume} {17}},\
  \bibinfo {pages} {471} (\bibinfo {year} {1998}{\natexlab{b}})}\BibitemShut
  {NoStop}%
\bibitem [{\citenamefont {Makihara}\ \emph {et~al.}(2002)\citenamefont
  {Makihara}, \citenamefont {Kida},\ and\ \citenamefont
  {Miura}}]{Makihara2002}%
  \BibitemOpen
  \bibfield  {author} {\bibinfo {author} {\bibfnamefont {T.}~\bibnamefont
  {Makihara}}, \bibinfo {author} {\bibfnamefont {S.}~\bibnamefont {Kida}},\
  and\ \bibinfo {author} {\bibfnamefont {H.}~\bibnamefont {Miura}},\ }\bibfield
   {title} {\bibinfo {title} {Automatic tracking of low-pressure vortex},\
  }\href@noop {} {\bibfield  {journal} {\bibinfo  {journal} {J. Phys. Soc.
  Jpn.}\ }\textbf {\bibinfo {volume} {71}},\ \bibinfo {pages} {1622} (\bibinfo
  {year} {2002})}\BibitemShut {NoStop}%
\bibitem [{\citenamefont {Miura}(2002)}]{Miura2002}%
  \BibitemOpen
  \bibfield  {author} {\bibinfo {author} {\bibfnamefont {H.}~\bibnamefont
  {Miura}},\ }\bibfield  {title} {\bibinfo {title} {Analysis of vortex
  structures in compressible isotropic turbulence},\ }\href@noop {} {\bibfield
  {journal} {\bibinfo  {journal} {Comput. Phys. Commun.}\ }\textbf {\bibinfo
  {volume} {147}},\ \bibinfo {pages} {552} (\bibinfo {year}
  {2002})}\BibitemShut {NoStop}%
\bibitem [{\citenamefont {Miura}(2004)}]{Miura2004}%
  \BibitemOpen
  \bibfield  {author} {\bibinfo {author} {\bibfnamefont {H.}~\bibnamefont
  {Miura}},\ }\bibfield  {title} {\bibinfo {title} {Excitations of vortex waves
  in weakly compressible isotropic turbulence},\ }\bibfield  {journal}
  {\bibinfo  {journal} {J. Turbul.}\ }\textbf {\bibinfo {volume} {5}},\ \href
  {https://doi.org/10.1088/1468-5248/5/1/010} {10.1088/1468-5248/5/1/010}
  (\bibinfo {year} {2004})\BibitemShut {NoStop}%
\bibitem [{\citenamefont {Kawahara}(2005)}]{Kawahara2005}%
  \BibitemOpen
  \bibfield  {author} {\bibinfo {author} {\bibfnamefont {G.}~\bibnamefont
  {Kawahara}},\ }\bibfield  {title} {\bibinfo {title} {Energy dissipation in
  spiral vortex layers wrapped around a straight vortex tube},\ }\href@noop {}
  {\bibfield  {journal} {\bibinfo  {journal} {Phys. Fluids}\ }\textbf {\bibinfo
  {volume} {17}},\ \bibinfo {pages} {055111} (\bibinfo {year}
  {2005})}\BibitemShut {NoStop}%
\bibitem [{\citenamefont {Goto}()}]{Goto2009}%
  \BibitemOpen
  \bibfield  {author} {\bibinfo {author} {\bibfnamefont {S.}~\bibnamefont
  {Goto}},\ }\bibfield  {title} {\bibinfo {title} {Turbulent energy cascade
  caused by vortex stretching},\ }in\ \href@noop {} {\emph {\bibinfo
  {booktitle} {Advances in Turbulence XII (ed. B. Eckhardt)}}}\BibitemShut
  {NoStop}%
\bibitem [{\citenamefont {Oka}\ and\ \citenamefont {Goto}(2021)}]{Oka2021}%
  \BibitemOpen
  \bibfield  {author} {\bibinfo {author} {\bibfnamefont {S.}~\bibnamefont
  {Oka}}\ and\ \bibinfo {author} {\bibfnamefont {S.}~\bibnamefont {Goto}},\
  }\bibfield  {title} {\bibinfo {title} {Generalized sweep-stick mechanism of
  inertial-particle clustering in turbulence},\ }\href@noop {} {\bibfield
  {journal} {\bibinfo  {journal} {Phys. Rev. Fluids}\ }\textbf {\bibinfo
  {volume} {6}},\ \bibinfo {pages} {044605} (\bibinfo {year}
  {2021})}\BibitemShut {NoStop}%
\bibitem [{\citenamefont {Matsuura}\ and\ \citenamefont
  {Fukumoto}(2022)}]{Matsuura2022}%
  \BibitemOpen
  \bibfield  {author} {\bibinfo {author} {\bibfnamefont {K.}~\bibnamefont
  {Matsuura}}\ and\ \bibinfo {author} {\bibfnamefont {Y.}~\bibnamefont
  {Fukumoto}},\ }\bibfield  {title} {\bibinfo {title} {Hierarchical clustering
  method of volumetric vortical regions with application to the late stage of
  laminar-turbulent transition},\ }\href@noop {} {\bibfield  {journal}
  {\bibinfo  {journal} {Phys. Rev. Fluids}\ }\textbf {\bibinfo {volume} {7}},\
  \bibinfo {pages} {054703} (\bibinfo {year} {2022})}\BibitemShut {NoStop}%
\bibitem [{\citenamefont {Yoshida}\ and\ \citenamefont
  {Arimitsu}(2006)}]{Yoshida2006}%
  \BibitemOpen
  \bibfield  {author} {\bibinfo {author} {\bibfnamefont {K.}~\bibnamefont
  {Yoshida}}\ and\ \bibinfo {author} {\bibfnamefont {T.}~\bibnamefont
  {Arimitsu}},\ }\bibfield  {title} {\bibinfo {title} {Energy spectra in
  quantum fluid turbulence},\ }\href@noop {} {\bibfield  {journal} {\bibinfo
  {journal} {J. Low Temp. Phys.}\ }\textbf {\bibinfo {volume} {145}},\ \bibinfo
  {pages} {219} (\bibinfo {year} {2006})}\BibitemShut {NoStop}%
\bibitem [{\citenamefont {Gross}(1961)}]{Gross1961}%
  \BibitemOpen
  \bibfield  {author} {\bibinfo {author} {\bibfnamefont {E.~P.}\ \bibnamefont
  {Gross}},\ }\bibfield  {title} {\bibinfo {title} {Structure of a quantized
  vortex in boson systems},\ }\href@noop {} {\bibfield  {journal} {\bibinfo
  {journal} {Il Nuovo Cimento (1955-1965)}\ }\textbf {\bibinfo {volume} {20}},\
  \bibinfo {pages} {454} (\bibinfo {year} {1961})}\BibitemShut {NoStop}%
\bibitem [{\citenamefont {Pitaevskii}(1961)}]{Pitaevskii1961}%
  \BibitemOpen
  \bibfield  {author} {\bibinfo {author} {\bibfnamefont {L.~P.}\ \bibnamefont
  {Pitaevskii}},\ }\bibfield  {title} {\bibinfo {title} {Vortex lines in an
  imperfect bose gas},\ }\href@noop {} {\bibfield  {journal} {\bibinfo
  {journal} {Sov. Phys. JETP}\ }\textbf {\bibinfo {volume} {13}},\ \bibinfo
  {pages} {451} (\bibinfo {year} {1961})}\BibitemShut {NoStop}%
\bibitem [{\citenamefont {S.~Shukla}\ and\ \citenamefont
  {Pandit}()}]{Shukla2023}%
  \BibitemOpen
  \bibfield  {author} {\bibinfo {author} {\bibfnamefont {V.~S. A.~B.}\
  \bibnamefont {S.~Shukla}, \bibfnamefont {A.~K.~Verma}}\ and\ \bibinfo
  {author} {\bibfnamefont {R.}~\bibnamefont {Pandit}},\ }\bibfield  {title}
  {\bibinfo {title} {Inertial particles in superfluid turbulence: Coflow and
  counterflow},\ }\href@noop {} {\bibfield  {journal} {\bibinfo  {journal}
  {Phys. Fluids}\ }\textbf {\bibinfo {volume} {35}}}\BibitemShut {NoStop}%
\bibitem [{\citenamefont {Sasa}\ \emph {et~al.}(2011)\citenamefont {Sasa},
  \citenamefont {Kano}, \citenamefont {Machida}, \citenamefont {L'vov},
  \citenamefont {Rudenko},\ and\ \citenamefont {Tsubota}}]{Sasa2011}%
  \BibitemOpen
  \bibfield  {author} {\bibinfo {author} {\bibfnamefont {N.}~\bibnamefont
  {Sasa}}, \bibinfo {author} {\bibfnamefont {T.}~\bibnamefont {Kano}}, \bibinfo
  {author} {\bibfnamefont {M.}~\bibnamefont {Machida}}, \bibinfo {author}
  {\bibfnamefont {V.~S.}\ \bibnamefont {L'vov}}, \bibinfo {author}
  {\bibfnamefont {O.}~\bibnamefont {Rudenko}},\ and\ \bibinfo {author}
  {\bibfnamefont {M.}~\bibnamefont {Tsubota}},\ }\bibfield  {title} {\bibinfo
  {title} {Energy spectra of quantum turbulence: Large-scale simulation and
  modeling},\ }\href@noop {} {\bibfield  {journal} {\bibinfo  {journal} {Phys.
  Rev. B}\ }\textbf {\bibinfo {volume} {84}},\ \bibinfo {pages} {054525}
  (\bibinfo {year} {2011})}\BibitemShut {NoStop}%
\bibitem [{\citenamefont {Baggaley}(2012)}]{Baggaley2012}%
  \BibitemOpen
  \bibfield  {author} {\bibinfo {author} {\bibfnamefont {A.~W.}\ \bibnamefont
  {Baggaley}},\ }\bibfield  {title} {\bibinfo {title} {The importance of vortex
  bundles in quantum turbulence at absolute zero},\ }\href@noop {} {\bibfield
  {journal} {\bibinfo  {journal} {Phys. Fluids}\ }\textbf {\bibinfo {volume}
  {24}},\ \bibinfo {pages} {055109} (\bibinfo {year} {2012})}\BibitemShut
  {NoStop}%
\bibitem [{\citenamefont {Rorai}\ \emph {et~al.}(2016)\citenamefont {Rorai},
  \citenamefont {Skipper}, \citenamefont {Kerr},\ and\ \citenamefont
  {Sreenivasan}}]{Rorai2016}%
  \BibitemOpen
  \bibfield  {author} {\bibinfo {author} {\bibfnamefont {C.}~\bibnamefont
  {Rorai}}, \bibinfo {author} {\bibfnamefont {J.}~\bibnamefont {Skipper}},
  \bibinfo {author} {\bibfnamefont {R.~M.}\ \bibnamefont {Kerr}},\ and\
  \bibinfo {author} {\bibfnamefont {K.~R.}\ \bibnamefont {Sreenivasan}},\
  }\bibfield  {title} {\bibinfo {title} {Approach and separation of quantum
  vortices with balanced cores},\ }\href@noop {} {\bibfield  {journal}
  {\bibinfo  {journal} {Journal of Fluid Mecnanics}\ }\textbf {\bibinfo
  {volume} {808}},\ \bibinfo {pages} {055301(R)} (\bibinfo {year}
  {2016})}\BibitemShut {NoStop}%
\bibitem [{\citenamefont {Krstulovic}\ and\ \citenamefont
  {Brachet}(2011)}]{Krstulovic2011}%
  \BibitemOpen
  \bibfield  {author} {\bibinfo {author} {\bibfnamefont {G.}~\bibnamefont
  {Krstulovic}}\ and\ \bibinfo {author} {\bibfnamefont {M.}~\bibnamefont
  {Brachet}},\ }\bibfield  {title} {\bibinfo {title} {Dispersive bottleneck
  delaying thermalization of turbulent bose-einstein condensates},\ }\href@noop
  {} {\bibfield  {journal} {\bibinfo  {journal} {Phys. Rev. Lett.}\ }\textbf
  {\bibinfo {volume} {106}},\ \bibinfo {pages} {115303} (\bibinfo {year}
  {2011})}\BibitemShut {NoStop}%
\bibitem [{\citenamefont {Lvov}\ and\ \citenamefont
  {Nazarenko}(2010)}]{Lvov2010}%
  \BibitemOpen
  \bibfield  {author} {\bibinfo {author} {\bibfnamefont {V.~S.}\ \bibnamefont
  {Lvov}}\ and\ \bibinfo {author} {\bibfnamefont {S.}~\bibnamefont
  {Nazarenko}},\ }\bibfield  {title} {\bibinfo {title} {Spectrum of kelvin-wave
  turbulence in superfluids},\ }\href@noop {} {\bibfield  {journal} {\bibinfo
  {journal} {JETP Letters}\ }\textbf {\bibinfo {volume} {91}},\ \bibinfo
  {pages} {428} (\bibinfo {year} {2010})}\BibitemShut {NoStop}%
\bibitem [{\citenamefont {G\"{u}nter}\ and\ \citenamefont
  {Theisel}(2018)}]{Gunter2018}%
  \BibitemOpen
  \bibfield  {author} {\bibinfo {author} {\bibfnamefont {T.}~\bibnamefont
  {G\"{u}nter}}\ and\ \bibinfo {author} {\bibfnamefont {H.}~\bibnamefont
  {Theisel}},\ }\bibfield  {title} {\bibinfo {title} {The state of the art in
  vortex extraction},\ }\href@noop {} {\bibfield  {journal} {\bibinfo
  {journal} {Comput Graph Forum}\ }\textbf {\bibinfo {volume} {37}},\ \bibinfo
  {pages} {149} (\bibinfo {year} {2018})}\BibitemShut {NoStop}%
\bibitem [{\citenamefont {Yao}\ and\ \citenamefont
  {Hussain}(2021)}]{YaoHussain2021}%
  \BibitemOpen
  \bibfield  {author} {\bibinfo {author} {\bibfnamefont {J.}~\bibnamefont
  {Yao}}\ and\ \bibinfo {author} {\bibfnamefont {F.}~\bibnamefont {Hussain}},\
  }\bibfield  {title} {\bibinfo {title} {Polarized vortex reconnection},\
  }\href {https://doi.org/10.1017/jfm.2021.517} {\bibfield  {journal} {\bibinfo
   {journal} {J. Fluid Mech.}\ }\textbf {\bibinfo {volume} {922}},\ \bibinfo
  {pages} {A19} (\bibinfo {year} {2021})}\BibitemShut {NoStop}%
\bibitem [{\citenamefont {Nakayama}\ and\ \citenamefont
  {Mizushima}(2017)}]{Nakayama2017}%
  \BibitemOpen
  \bibfield  {author} {\bibinfo {author} {\bibfnamefont {K.}~\bibnamefont
  {Nakayama}}\ and\ \bibinfo {author} {\bibfnamefont {L.~D.}\ \bibnamefont
  {Mizushima}},\ }\bibfield  {title} {\bibinfo {title} {A numerical analysis
  for identification of flow transition in vortex generation in terms of local
  flow topology},\ }\href {https://doi.org/10.1299/jfst.2017jfst0027}
  {\bibfield  {journal} {\bibinfo  {journal} {J. Fluid Sci. Technol.}\ }\textbf
  {\bibinfo {volume} {12}},\ \bibinfo {pages} {JFST0027} (\bibinfo {year}
  {2017})}\BibitemShut {NoStop}%
\bibitem [{\citenamefont {Goto}\ and\ \citenamefont {Kida}(2003)}]{Goto2003}%
  \BibitemOpen
  \bibfield  {author} {\bibinfo {author} {\bibfnamefont {S.}~\bibnamefont
  {Goto}}\ and\ \bibinfo {author} {\bibfnamefont {S.}~\bibnamefont {Kida}},\
  }\bibfield  {title} {\bibinfo {title} {Enhanced stretching of material lines
  by antiparallel vortex pairs in turbulence},\ }\href@noop {} {\bibfield
  {journal} {\bibinfo  {journal} {Fluid Dyn. Res.}\ }\textbf {\bibinfo {volume}
  {33}},\ \bibinfo {pages} {403} (\bibinfo {year} {2003})}\BibitemShut
  {NoStop}%
\end{thebibliography}
%

\end{document}